\documentclass{article}

\PassOptionsToPackage{numbers, compress}{natbib}

\usepackage[main, final]{neurips_2026}

\usepackage[utf8]{inputenc} 
\usepackage[T1]{fontenc}    
\usepackage{hyperref}       
\usepackage{url}            
\usepackage{booktabs}       
\usepackage{amsfonts}       
\usepackage{nicefrac}       
\usepackage{microtype}      
\usepackage{xcolor}         
\usepackage{enumitem}
\usepackage{multirow}
\usepackage{kotex}
\usepackage{mathtools}
\usepackage{booktabs}
\usepackage{makecell}
\usepackage{pifont}
\usepackage{tabularx}
\usepackage[normalem]{ulem}
\newcommand{\proposed}{\textsf{Self-EvolveRec}}

\usepackage{subcaption}
\usepackage{wrapfig}

\title{Self-EvolveRec: Self-Evolving Recommender Systems with LLM-based Directional Feedback}

\author{%
  \textbf{Sein Kim}\textsuperscript{1},
  \textbf{Sangwu Park}\textsuperscript{1},
  \textbf{Hongseok Kang}\textsuperscript{1},
  \textbf{Wonjoong Kim}\textsuperscript{1},
  \textbf{Jimin Seo}\textsuperscript{1} \\
  \textbf{Yeonjun In}\textsuperscript{1},
  \textbf{Kanghoon Yoon}\textsuperscript{1},
  \textbf{Hyunsik Jeon}\textsuperscript{2},
  \textbf{Chanyoung Park}\textsuperscript{1,*} \\[0.4em]
  \textsuperscript{1}KAIST
  \qquad
  \textsuperscript{2}Roblox \\[0.3em]
  \texttt{\{rlatpdlsgns, sangwu.park, ghdtjr0311, wjkim,}\\
  \texttt{jimin.seo, yeonjun.in, ykhoon08, cy.park\}@kaist.ac.kr}\\
  \texttt{hjeon@roblox.com}\\[0.2em]
  {\small \textsuperscript{*}Corresponding author}
}

\begin{document}

\maketitle

\vspace{-2ex}
\begin{abstract}
Traditional methods for automating recommender system design, such as Neural Architecture Search (NAS), are often constrained by a fixed search space defined by human priors, limiting innovation to pre-defined operators. While recent LLM-driven code evolution frameworks shift fixed search space target to open-ended program spaces, they primarily rely on scalar metrics (e.g., NDCG, Hit Ratio) that fail to provide qualitative insights into model failures or directional guidance for improvement. To address this, we propose~\proposed, a novel framework that establishes a directional feedback loop by integrating a User Simulator for qualitative critiques and a Model Diagnosis Tool for quantitative internal verification. Furthermore, we introduce a Diagnosis Tool - Model Co-Evolution strategy to ensure that evaluation criteria dynamically adapt as the recommendation architecture evolves. Extensive experiments demonstrate that~\proposed~significantly outperforms state-of-the-art NAS and LLM-driven code evolution baselines in both recommendation performance and user satisfaction. Our code is available at \url{https://github.com/Sein-Kim/self_evolverec}.
\end{abstract}

\section{Introduction}
\vspace{-1ex}
Driven by the growth of online data~\cite{10.1145/2959100.2959190, 10.1145/2843948}, recommender systems have evolved from Matrix Factorization~\cite{he2017neural, 10.5555/2981562.2981720, 10.1145/2959100.2959165, 10.1145/3726302.3729953} to Graph Neural Networks~\cite{wang2019neural, he2020lightgcn} and Transformers~\cite{kang2018self,sun2019bert4rec}. However, optimal performance relies heavily on the entire recommendation pipeline (e.g., loss functions, negative sampling) rather than model architecture alone~\cite{naumov2019deep, zou2025survey}, making manual refinement of this pipeline by human expertise inefficient and costly~\cite{zheng2023automl}.
To mitigate manual design inefficiencies, Automated Machine Learning (AutoML) techniques, such as Neural Architecture Search (NAS)~\cite{zoph2017neural, elsken2019neural} have emerged as a prominent approach to automating the discovery of optimal recommendation architectures~\cite{liu2020autofis,song2020towards,krishna2021differentiable,zhang2023nasrec,li2022autolossgen}.

However, these methodologies are inherently constrained by a fixed search space bounded by human priors~\cite{ci2021evolving, real2020automl}, limiting optimization to symbolic combinations within a closed pool of pre-defined operators. Due to this lack of generative expressivity, existing NAS methods struggle to address non-architectural components—such as loss functions and data processing—and fail to jointly optimize the entire recommendation pipeline. Consequently, achieving comprehensive system optimization requires shifting from a closed operator pool to an open-ended program space~\cite{real2020automl}. 

To realize open-ended optimization, a new paradigm known as LLM-driven code evolution has emerged. Pioneering frameworks such as FunSearch~\cite{romera2024mathematical} and Eureka~\cite{ma2023eureka} validated LLM-driven code evolution approach by leveraging LLMs to optimize isolated functions or specific logic components. Substantially expanding this scope, AlphaEvolve~\cite{novikov2025alphaevolve} targets entire codebases rather than single functions. By orchestrating an autonomous evolutionary pipeline, AlphaEvolve iteratively optimizes complex algorithmic structures and computational stacks through direct code modifications. Complementing AlphaEvolve, DeepEvolve~\cite{liu2025scientific} extends the evolution loop by incorporating Retrieval-Augmented Generation (RAG)~\cite{lewis2020retrieval, gao2023retrieval}. By retrieving academic papers from arXiv, it leverages external scientific knowledge to facilitate systematic idea generation and codebase refinement. Unlike NAS confined to combining existing modules, these LLM-driven approaches enable open-ended optimization akin to human researchers, allowing the invention of novel components beyond pre-defined design space.


Despite this open-ended capability, existing LLM-driven methods are guided primarily by scalar metrics (e.g., accuracy, MSE) that lack diagnostic insight, restricting evolution to an undirected trial-and-error search. This is particularly problematic in recommender systems, where—unlike mathematical problems with deterministic ground-truth~\cite{novikov2025alphaevolve}—failures are multifaceted: a drop in NDCG or Hit Ratio cannot reveal whether the cause is excessive popularity bias, insufficient category diversity, or a failure to capture short-term interests~\cite{10.1145/3240323.3240372,10.1145/1060745.1060754}. Thus, effective code evolution in this domain requires \textit{directional feedback} that diagnoses root causes of failure and reflects user experience to guide the evolution process.

To address this challenge, we propose~\proposed, an LLM-driven code evolution framework that orchestrates a directional feedback loop by integrating a \textbf{User Simulator} (\text{SIM}) and a \textbf{Model Diagnosis Tool} (\text{DIAG}). \text{SIM} evaluates recommendation lists through diverse user personas, providing qualitative natural language critiques (e.g., \textit{"I seek low-cost accessories, not expensive electronics"}) that explain \textit{why} a recommendation fails—an insight scalar metrics cannot offer. To complement \text{SIM} and detect deficiencies invisible to user perception (e.g., embedding collapse), \text{DIAG} quantitatively probes the model's structural properties, such as the geometric distribution of item representations. By corroborating qualitative critiques with quantitative diagnostics,~\proposed~accurately pinpoints the root causes of failure within the recommendation pipeline.

However, this static design becomes inadequate as the pipeline evolves: a fixed \text{DIAG} cannot inspect newly introduced components (e.g., multi-modal encoders, novel loss functions) and lacks metrics to verify newly emerging \text{SIM} critiques (e.g., "lack of diversity"). To address this, we design \textit{Diagnosis Tool - Model Co-Evolution}, which dynamically updates \text{DIAG} to (i) inspect new components and (ii) translate \text{SIM}'s qualitative critiques into measurable metrics. We additionally introduce an optional efficient variant based on zero-cost proxies~\cite{lee2024az,abdelfattah2021zero} for scalable deployment. Extensive experiments demonstrate that~\proposed~outperforms NAS and LLM-driven code evolution baselines, with directional feedback yielding substantial gains in recommendation performance, user satisfaction, and the technical quality of the evolved algorithmic logic. We defer a detailed discussion of related work to App.~\ref{app related work}.

\begin{figure*}[t]
    \centering    
    \vspace{-1ex}\includegraphics[width=0.95\linewidth]{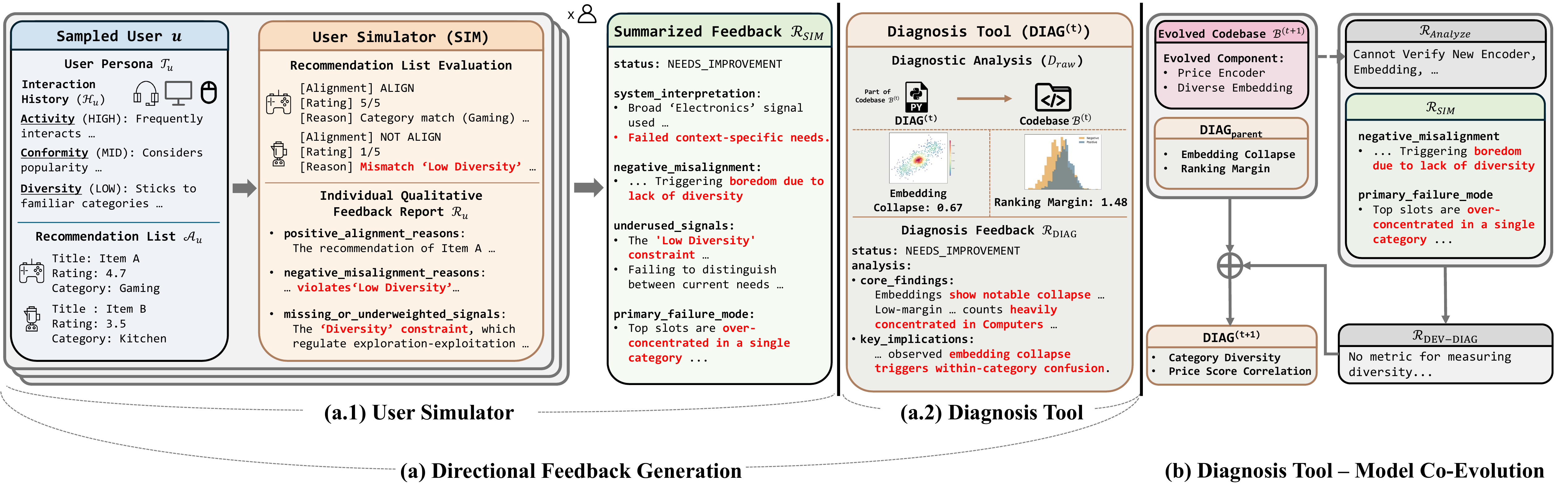}
    \vspace{-2ex}
    \captionsetup{font=small}
    \caption
    {Two core mechanism of~\proposed. (a) is the overview of the Directional Feedback Generation: (a.1) is the user simulator, (a.2) is the model diagnosis tool. (b) is the Diagnosis Tool - Model Co-evolution.}
    \vspace{-3ex}
    \label{fig: directional_feedback}
\end{figure*}

\section{Problem Definition}
\label{Sec Problem Definition}
\vspace{-2ex}
\noindent\textbf{Dataset.}
Let $\mathcal{D} = (\mathcal{U}, \mathcal{V}, \mathcal{E}, \mathcal{X})$ denote the dataset with user set $\mathcal{U}$ and item set $\mathcal{V}$.
The interaction set $(u, v, t_{u,v}, r_{u,v}, \text{rev}_{u,v}) \in \mathcal{E}$ indicates that user $u \in \mathcal{U}$ interacted with item $v \in \mathcal{V}$ at timestamp $t_{u,v}$, providing a rating $r_{u,v}$ and a textual review $\text{rev}_{u,v}$. Based on $\mathcal{E}$, we define the interaction history for user $u$ as $\mathcal{H}_u = \{v \mid (u, v, t_{u,v}) \in \mathcal{E}\}$, chronologically ordered by $t_{u,v}$. Additionally, $x_v \in \mathcal{X}$ denotes the set of side information, where $x_v$ represents the raw attribute set for item $v$ (e.g., category, title, and price).

\smallskip
\noindent\textbf{Optimization Goal.}
Let $\mathcal{B} \in \mathbb{S}$ denote the entire codebase governing the \textit{recommendation pipeline} within the \textbf{open-ended program space} $\mathbb{S}$. The evolutionary process starts with a \textit{seed codebase} $\mathcal{B}^{(0)}$ and proceeds through $T$ iterations, where $\mathcal{B}^{(t)}$ represents the evolved codebase at iteration $t$. Specifically, $\mathcal{B}^{(0)}$ constitutes a fully functional seed recommendation pipeline, encapsulating a seed recommender architecture $f_{\mathcal{B}^{(0)}}(\cdot ; \theta_{\mathcal{B}^{(0)}})$ (e.g., NCF~\cite{he2017neural}), data processing logic (e.g., basic loaders for interactions $\mathcal{E}$ and raw attributes $\mathcal{X}$), and a standard optimization loop (e.g., loss computation, and parameter updates). Our goal is to evolve this seed codebase $\mathcal{B}^{(0)}$ into an optimal codebase $\mathcal{B}^*$ that maximizes a standard recommendation metric $\mathcal{M}$ (e.g., Hit Ratio, NDCG) within $T$ iterations. We formulate this code evolution task as a bi-level optimization problem:

\begin{equation}
\small
\begin{aligned}
    \mathcal{B}^* = \operatorname*{argmax}_{\mathcal{B} \in \mathbb{S}} \underbrace{\mathcal{M} \left( f_{\mathcal{B}} (\mathcal{E}_{\text{val}}, \mathcal{X} ; \theta_{\mathcal{B}}^*) \right)}_{\textsf{score}(\mathcal{B})} \quad 
    \text{s.t.} \,\, \theta_{\mathcal{B}}^* = \operatorname*{argmin}_{\theta} \mathcal{L}_{\mathcal{B}} \left( f_{\mathcal{B}} (\mathcal{E}_{\text{train}}, \mathcal{X} ; \theta) \right)
\end{aligned}
\label{Eq Optimization Problem}
\end{equation}
where $\mathcal{L}_{\mathcal{B}}$ represents the loss function defined within $\mathcal{B}$. Here, $\theta_{\mathcal{B}}^*$ indicates the optimal model parameters for the architecture defined by $\mathcal{B}$, learned on the training set $\mathcal{E}_{\text{train}}$ by minimizing $\mathcal{L}_{\mathcal{B}}$.

\section{Proposed Framework:~\proposed}
\label{Sec methodology}
In this section, we propose ~\proposed, a novel framework designed to enable LLMs to perform directional feedback-based evolutionary code optimization for recommender systems.
As illustrated in Figure~\ref{fig: directional_feedback},~\proposed~operates through two core mechanisms to overcome the limitations of existing metric-based approaches~\cite{novikov2025alphaevolve, liu2025scientific}. \textbf{First}, we establish \textit{Directional Feedback Generation} (Sec.~\ref{Sec Directional Feedback}, Figure~\ref{fig: directional_feedback} (a)), which integrates a \textit{User Simulator} (Sec.~\ref{Sec User Simulator}) and a \textit{Model Diagnosis Tool} (Sec.~\ref{Sec Diagnosis Tool}) to provide qualitative and quantitative guidance. 
This feedback then guides the \textit{Evolution Pipeline} (Sec.~\ref{Sec Evolution Pipeline}, Figure~\ref{fig: framework}) to perform precise, directed code modifications.
\textbf{Second}, we introduce \textit{Diagnosis Tool - Model Co-evolution} (Sec.~\ref{Sec Co-evolution}, Figure~\ref{fig: directional_feedback} (b)), a strategy that ensures the diagnosis tool to dynamically adapt to the structural changes of the recommendation pipeline.

\subsection{Directional Feedback Generation}
\label{Sec Directional Feedback}
While scalar metrics (e.g., NDCG) quantify how well a recommendation pipeline performs, they lack the semantic depth to explain why it fails or how to resolve it. To address this, we introduce \textbf{Directional Feedback}, which integrates qualitative critiques from the User Simulator and quantitative verifications from the Model Diagnosis Tool. This mechanism translates non-interpretable numerical metrics into actionable insights, enabling the LLM to pinpoint and resolve structural deficiencies within the recommendation pipeline.

\subsubsection{\textbf{User Simulator: Qualitative Critique}}
\label{Sec User Simulator}
We employ a User Simulator (\text{SIM}) to complement standard metrics with qualitative directional feedback. While scalar metrics (e.g., NDCG) quantify \textit{how well} a model performs, they fail to reveal the root causes of failure~\cite{10.1145/1125451.1125659} such as insufficient category diversity or an inability to capture short-term interests. By adopting the agentic paradigm~\cite{zhang2024generative, ma2025pub}, as depicted in Figure~\ref{fig: directional_feedback}~(a.1), our simulator acts as a diverse set of virtual users, offering explicit natural language critiques that pinpoint specific deficiencies in the recommendation pipeline.


To ensure behavioral realism and heterogeneity, we characterize each simulated user $u$ through a structured persona $\mathcal{T}_u$, constructed by combining the user's interaction history $\mathcal{H}_u$ with sociopsychological traits~\cite{zhang2024generative,10.1145/2187980.2188230}. Following~\cite{zhang2024generative}, we use three traits—\textbf{Activity}, \textbf{Conformity}, and \textbf{Diversity}—capturing engagement level, mainstream adherence, and interest breadth, respectively, each quantized into three levels (LOW, MID, HIGH) and mapped to natural language descriptions (e.g., \textit{"Activity (HIGH): Frequently interacts with the recommender..."}); see App.~\ref{app persona} for details. Our framework is agnostic to specific persona definitions, and alternative schemes such as the Big Five~\cite{ma2025pub, Goldberg1992THEDO} can be integrated, as explored in App.~\ref{exp big five}.

The feedback generation process proceeds in two steps. First, \text{SIM} conducts an individual assessment for a sampled user $u$ by evaluating the recommendation list $\mathcal{A}_u$ from a trained recommender $f_{\mathcal{B}}$ (e.g., NCF) based on their persona $\mathcal{T}_u$ and interaction history $\mathcal{H}_u$. This assessment identifies specific behavioral or semantic misalignment, which is formalized into an individual qualitative feedback report $R_u = \text{LLM}(\mathcal{I}_{\text{SIM}}, \mathcal{T}_u, \mathcal{H}_u, \mathcal{A}_u)$, where $\mathcal{I}_{\text{SIM}}$ is instruction for user simulator. {Throughout the paper, we denote the task-specific instructions guiding the LLM-agent as $\mathcal{I}_{\text{task}}$ (e.g., $\mathcal{I}_{\text{SIM}}$, $\mathcal{I}_{\text{PLAN}}$, $\mathcal{I}_{\text{CODE}}$). While we describe the high-level objective of each instruction within the main text, the exact prompt templates are provided in App.~\ref{app prompts task}.}
Second, to mitigate individual user bias and capture common failure patterns, we summarize reports from a set of sampled users $\mathcal{U}_{\text{sample}}$ into a comprehensive summary $\mathcal{R}_{\text{SIM}} = \text{LLM}(\mathcal{I}_{\text{SUMMARIZE}}, \{R_u \mid u \in \mathcal{U}_{\text{sample}}\})$,
where $\mathcal{I}_{\text{SUMMARIZE}}$ guides the LLM to abstract common failure patterns from individual critiques.

\subsubsection{\textbf{Model Diagnosis Tool: Quantitative Verification}}
\label{Sec Diagnosis Tool}
While the \text{SIM} generates qualitative feedback from user experience, relying solely on this feedback fails to detect hidden structural issues in terms of the model, such as embedding collapse. To address this, we introduce the Model Diagnosis Tool ($\text{DIAG}$; denoted as $\text{DIAG}^{(t)}$ at iteration $t$), which is a computational probing module designed to verify structural or behavioral deficiencies of the recommender system. As depicted in Figure~\ref{fig: directional_feedback}~(a.2), unlike the \text{SIM},  $\text{DIAG}^{(t)}$ directly accesses the model parameters $\theta$ and data loaders within the codebase $\mathcal{B}^{(t)}$ to conduct a systematic validity check.

As an initial instantiation of the diagnosis tool  $\text{DIAG}^{(0)} \in \mathcal{B}^{(0)}$, we implement two foundational probes designed to detect common structural failures in recommender models:

\smallskip
\noindent\textbf{1) Embedding Collapse:} 
To detect a state where representations degenerate into a narrow subspace losing discriminative power, $\text{DIAG}$ computes the mean pairwise cosine similarity across sampled item embeddings. A high similarity score serves as a proxy for representation degeneration.

\smallskip
\noindent\textbf{2) Ranking Margin:} {$\text{DIAG}$ evaluates decision boundaries by analyzing the ranking margin $\Delta_{u,v} = s(u,v) - s(u, v')$ for all users $u \in \mathcal{U}$ and their observed interactions $v \in \mathcal{H}_u$. Here, $s(u, v)$ and $s(u, v')$ denote the predicted logits for the ground-truth item $v$ and a randomly sampled negative item $v' \notin \mathcal{H}_u$, respectively. To assess overall discriminative power, $\text{DIAG}$ computes the global average margin $\mathbb{E}_{u \in \mathcal{U}, v \in \mathcal{H}_u} [\Delta_{u,v}]$. Specifically, a high margin indicates robust discrimination between ground-truth item and negative item, whereas a low or negative margin indicates a failure to distinguish ground-truth. To pinpoint the potential failure modes of such failures, \text{DIAG} aggregates cases with extremely low margins and counts their common attributes (e.g., specific categories like \textit{Computers} in Figure~\ref{fig: directional_feedback} (a.2) - "core\_findings").}
\textcolor{blue}{}

These probes generate a set of raw numerical diagnostics analysis $D_{\text{raw}} = \text{DIAG}^{(t)}(\mathcal{B}^{(t)})$. Subsequently, to bridge the gap between numerical diagnostics and algorithmic solutions, an LLM acts as a senior researcher to interpret these signals, converting them into a structured diagnosis report $\mathcal{R}_{\text{DIAG}} = \text{LLM}(\mathcal{I}_{\text{DIAG}}, D_{\text{raw}})$.
Furthermore, $\text{DIAG}$ serves as a verification mechanism to check whether the qualitative deficiencies pointed out by $\text{SIM}$ are actually resolved (detailed in Sec.~\ref{Sec Co-evolution}).


\vspace{-0.5ex}
\subsection{Evolution Pipeline}
\label{Sec Evolution Pipeline}
\vspace{-1ex}
\begin{wrapfigure}{r}{0.31\textwidth}
  \begin{center}
    \vspace{-2ex}
    \includegraphics[width=0.31\textwidth]{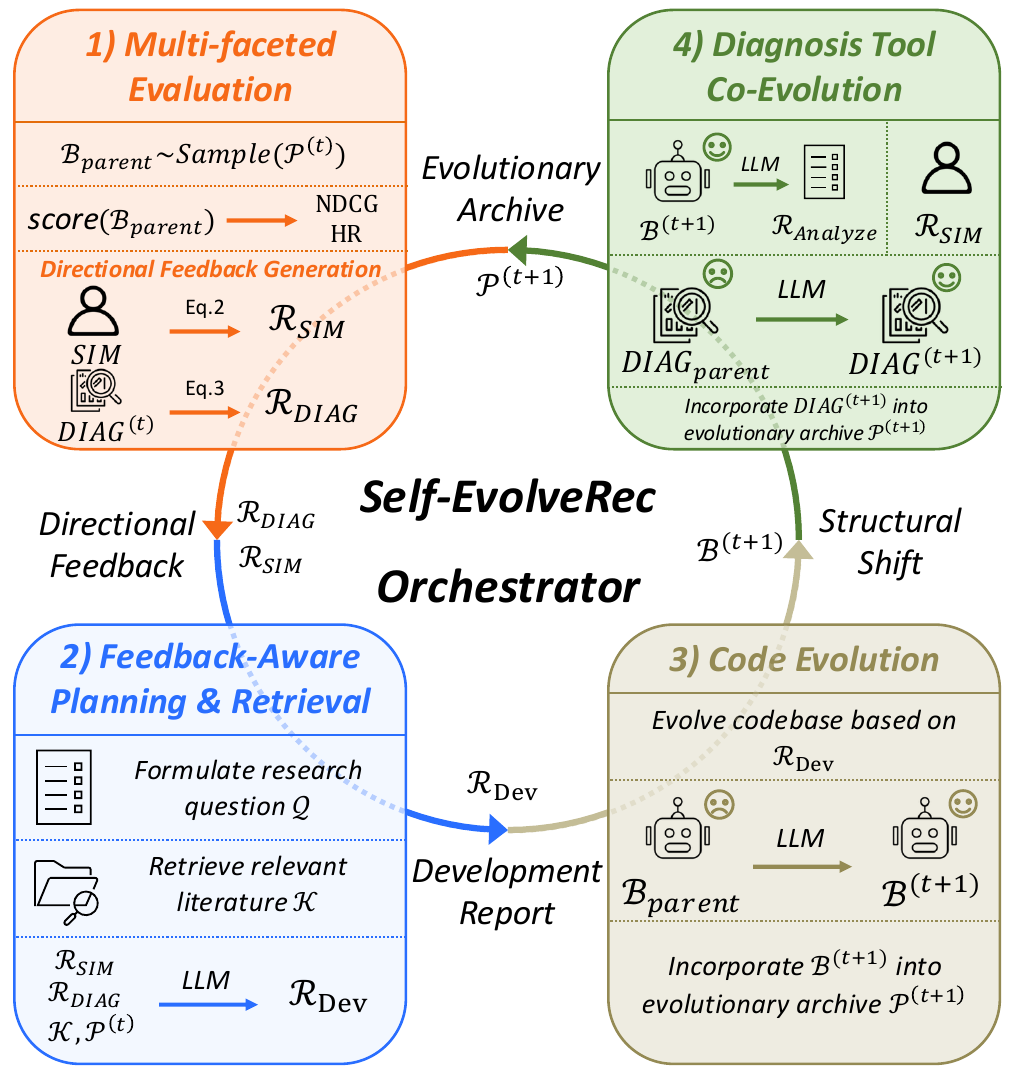}
    \vspace{-10pt}
    \caption{Overall evolutionary pipeline of \proposed.}
    \label{fig: framework}
    \vspace{-2ex}
  \end{center}
\end{wrapfigure}

In this section, we detail the iterative execution workflow of \proposed, designed to autonomously refine the codebase $\mathcal{B}$ through cycles of evaluation, reasoning, and evolution.
To support these cycles, \proposed~adopts a population-based exploration strategy~\cite{10.5555/915973, liu2025scientific, romera2024mathematical, novikov2025alphaevolve} by maintaining an Evolutionary Archive $\mathcal{P}^{(t)}$ that stores the comprehensive history of codebases and feedback. At each iteration, a target codebase is selected from this archive to serve as the parent, denoted as $\mathcal{B}_{\text{parent}}\sim\text{Sample}(\mathcal{P}^{(t)})$.
{Unlike existing studies~\cite{novikov2025alphaevolve,liu2025scientific} that rely solely on scalar metrics, which often lead to aimless exploration,~\proposed~actively integrates the \textbf{Directional Feedback} mechanisms (Sec.~\ref{Sec Directional Feedback}).} By integrating qualitative insights from the \textit{user feedback report} ($\mathcal{R}_{\text{SIM}}$) and quantitative findings from the \textit{diagnosis report} ($\mathcal{R}_{\text{DIAG}}$), \proposed~shifts the focus from mere scalar metrics to pinpoint the root causes of failure. As illustrated in Figure~\ref{fig: framework},~\proposed~follows a four-phase workflow to address these failures and evolve the codebase.

We first detail the three-phase workflow dedicated to refinement of the recommendation codebase, i.e., 1), 2), and 3) in Figure~\ref{fig: framework} ($\mathcal{B}_{\text{parent}} \rightarrow \mathcal{B}^{(t+1)}$),
followed by the Diagnosis Tool Co-Evolution stage, i.e., 4) in Figure~\ref{fig: framework} (Sec.~\ref{Sec Co-evolution}).

\smallskip
\noindent \textbf{1) Multi-faceted Evaluation.}
First, \proposed~evaluates the currently selected codebase $\mathcal{B}_{\text{parent}}$ to obtain the scalar score shown in Equation~\ref{Eq Optimization Problem} (i.e., \textsf{score}$(\mathcal{B}_{\text{parent}})$), while simultaneously generating the user feedback $\mathcal{R}_{\text{SIM}}$ and diagnosis report $\mathcal{R}_{\text{DIAG}}$\footnote{To maximize efficiency, if the selected codebase $\mathcal{B}_{\text{parent}}$ is already recorded in the archive $\mathcal{P}^{(t)}$ with complete evaluation logs, \proposed~retrieves the cached results instead of re-executing the evaluation process.}. These outputs serve as the directional context for the subsequent planning.

\smallskip
\noindent \textbf{2) Feedback-Aware Planning \& Retrieval.}
{\proposed~introduces a \textit{feedback-aware planning \& retrieval} that shifts the RAG methods in previous work~\cite{liu2025scientific} from general-purpose knowledge gathering to targeted failure resolution. By conditioning on directional feedback from $\mathcal{R}_{\text{SIM}}$ and $\mathcal{R}_{\text{DIAG}}$, the LLM agent acts as a planner to formulate precise feedback targeted research queries $\mathcal{Q} = \text{LLM}(\mathcal{I}_{\text{PLAN}}, \allowbreak \mathcal{R}_{\text{SIM}}, \allowbreak \mathcal{R}_{\text{DIAG}}, \allowbreak \mathcal{P}^{(t)})$.}
Conditioned on specific failure modes and the evolutionary archive $\mathcal{P}^{(t)}$, the agent generates targeted queries $\mathcal{Q}$ (e.g., "Retrieve methods to mitigate category mismatch identified in user reports") to retrieve relevant academic literature $\mathcal{K}$ from online sources, including arXiv\footnote{{We set the number of literature (i.e., $|\mathcal{K}|$) to 3. Refer to App.~\ref{app: literature sample} for experiments regarding the number of literature.}}. {These targeted queries ensure a retrieval focusing on resolving the identified structural deficiencies, rather than broad algorithmic exploration.}
These insights are then integrated into a structured \textit{Development Report} $\mathcal{R}_{\text{Dev}} = \text{LLM}(\mathcal{I}_{\text{REPORT}}, \mathcal{R}_{\text{SIM}}, \mathcal{R}_{\text{DIAG}}, \mathcal{P}^{(t)}, \mathcal{K})$,
where $\mathcal{R}_{\text{Dev}}$ outlines the algorithmic modifications required to address the identified issues.

\smallskip
\noindent \textbf{3) Code Evolution.}
Guided by $\mathcal{R}_{\text{Dev}}$, the agent implements code-level modifications to instantiate the updated codebase $\mathcal{B}^{(t+1)} = \text{LLM}(\mathcal{I}_{\text{CODE}}, \mathcal{R}_{\text{Dev}}, \mathcal{B}_{\text{parent}}, \mathcal{P}^{(t)})$.
Upon successful execution, the new codebase $\mathcal{B}^{(t+1)}$ is incorporated into the population history $\mathcal{P}^{(t+1)}$, following standard evolutionary protocols~\cite{novikov2025alphaevolve, liu2025scientific}. 
This cycle recursively refines $\mathcal{B}$ to optimize the objective defined in Equation~\ref{Eq Optimization Problem}.

\vspace{-1ex}
\subsection{Diagnosis Tool - Model Co-Evolution}
\label{Sec Co-evolution}
The evolution of the $\text{DIAG}$ (i.e., 4) in Figure~\ref{fig: framework}) is driven by two objectives:
\textbf{First}, as $\mathcal{B}$ undergoes structural transformations, such as the introduction of new loss functions or architectural layers, a static \text{DIAG} inevitably becomes obsolete.
$\text{DIAG}$ must continuously adapt to analyze and evaluate these new components effectively. \textbf{Second}, $\text{DIAG}$ serves to quantitatively verify the qualitative insights provided by the \text{SIM}.
While the \text{SIM} offers rich, human-like critiques that traditional metrics fail to capture (e.g., perceiving a recommendation list as "conceptually repetitive" despite high accuracy scores), these subjective critiques must be translated into measurable metrics to confirm their validity and assess their impact for precise code optimization.
Therefore, $\text{DIAG}$ evolves to stay compatible with the evolved codebase while formulating specific metrics that mathematically capture the essence of the \text{SIM}'s feedback. As depicted in Figure~\ref{fig: directional_feedback}~(b), if the \text{SIM} reports "boredom due to lack of diversity," $\text{DIAG}$ autonomously implements a tailored metric (e.g., measuring diversity among top-k items) to quantify this feedback precisely, ensuring that the subsequent evolution is guided by concrete objectives.

To address this, we implement a co-evolution mechanism that begins with understanding the structural shifts.
Since the codebase is continuously updated ($\mathcal{B}_{\text{parent}} \rightarrow \mathcal{B}^{(t+1)}$), the agent first scans the new codebase $\mathcal{B}^{(t+1)}$ and original $\text{DIAG}_{\text{parent}}$
to generate a structural analysis report $\mathcal{R}_{\text{Analyze}} = \text{LLM}(\mathcal{I}_{\text{Analyze}}, \mathcal{B}^{(t+1)}, \text{DIAG}_{\text{parent}})$.
The $\mathcal{R}_{\text{Analyze}}$ summarizes key information such as the updated execution flow, newly added modules, and modified loss functions. This blueprint enables the agent to design diagnostic criteria that are structurally compatible with the new architecture.

Subsequently, to synchronize the diagnosis tool with both the structural changes and the user feedback $\mathcal{R}_{\text{SIM}}$, the agent executes the evolution cycle. By integrating the qualitative $\mathcal{R}_{\text{SIM}}$ with the structural blueprint $\mathcal{R}_{\text{Analyze}}$, the agent identifies evaluation gaps (e.g., cannot verify new encoder, and embedding).
{Similar to the main pipeline (Sec.~\ref{Sec Evolution Pipeline}), it retrieves relevant methodologies $\mathcal{K}_{\text{DIAG}}$ using research queries $\mathcal{Q}_{\text{DIAG}} = \text{LLM}\left(\mathcal{I}_{\text{PLAN-DIAG}}, \mathcal{R}_{\text{SIM}}, \mathcal{R}_{\text{Analyze}}, \mathcal{P}^{(t)}\right)$ and then updates the \text{DIAG}} as:  $\text{DIAG}^{(t+1)} = \text{LLM}\left(\mathcal{I}_{\text{CODE-DIAG}},\mathcal{R}_{\text{Dev-DIAG}}, \mathcal{B}^{(t+1)}, \text{DIAG}_{\text{parent}}, \mathcal{P}^{(t)}\right)$, where $\mathcal{R}_{\text{Dev-DIAG}} = \text{LLM}\left(\mathcal{I}_{\text{REPORT-DIAG}},\mathcal{R}_{\text{SIM}}, \mathcal{R}_{\text{Analyze}}, \mathcal{K}_{\text{DIAG}}\right)$.



The new model diagnosis tool $\text{DIAG}^{(t+1)}$ is incorporated into the population history $\mathcal{P}^{(t+1)}$.
This adaptive process guarantees that $\text{DIAG}^{(t+1)}$ is equipped with both the logic to inspect new architectures and the specific metrics required to transform the simulator's qualitative feedback into actionable, quantitative signals. We further investigate the additional evolutionary pipeline of the User Simulator in experiments in Sec.~\ref{exp sim evolve}.

\subsection{Efficient Variant via Zero-Cost Proxies}
\label{Sec efficient variant}
While~\proposed~effectively guides evolution through directional feedback, fully training the recommender at every iteration is computationally expensive. To enable scalable deployment, we additionally introduce an \textbf{optional efficient variant} that replaces full training with zero-cost proxies~\cite{lee2024az,abdelfattah2021zero}, which estimate pipeline potential through a single forward-backward pass on a representative mini-batch. We design a composite proxy $\mathcal{M}_{\text{zero}}$ combining representation-health signals (effective ranks, top-1 singular value share) and gradient-health signals to surrogate standard recommendation metrics. Both DIAG and SIM are correspondingly adapted: DIAG leverages the sub-signals of $\mathcal{M}_{\text{zero}}$ as foundational probes, while SIM operates on a code-to-intent summary of the evolved codebase. Detailed formulations are provided in App.~\ref{app: zero cost detail}, and empirical validation showing up to 12$\times$ speedup with marginal performance degradation is provided in App.~\ref{app: zero cost eff}.

\vspace{-1ex}
\section{Experiments}
\label{sec: exp}
\vspace{-1ex}
\noindent\textbf{Datasets.} For evaluations, we used three Amzaon datasets~\cite{hou2024bridging} (CDs, Electronics, and Office) and the MovieLens dataset~\cite{10.1145/2827872}. Following prior works~\cite{kang2018self, sun2019bert4rec, kim2025image}, we use five-core datasets.
Detailed statistics for each dataset are provided in Table~\ref{app tab dataset} in App.~\ref{app dataset}.

\smallskip
\noindent\textbf{Baselines.}
We evaluate \proposed~against representative NAS-based recommender architecture search methods, i.e., AutoFIS~\cite{liu2020autofis}, NASRec~\cite{zhang2023nasrec} AutoLossGen~\cite{li2022autolossgen}, and DNS-Rec~\cite{zhang2024dns}, as well as recent LLM-driven evolutionary frameworks, i.e., AlphaEvolve~\cite{novikov2025alphaevolve} and DeepEvolve~\cite{liu2025scientific}.
Details regarding the baseline methods are provided in App.~\ref{app baseline}.

\smallskip
\noindent\textbf{Evaluation Protocol.} We use the leave-last-out strategy~\cite{kang2018self,sun2019bert4rec, tang2018personalized, 10.1145/3711896.3737035, 10.1145/3726302.3729959} for evaluation of recommender models, where we use the most recent item and second most item for testing and validation, respectively, and the remaining history for training.
Each test item is paired with 99 randomly sampled non-interacted items. We report performance using two standard metrics: Normalized Discounted Cumulative Gain (NDCG@5) and Hit Ratio (HR@5). Crucially, the standard ranking gains (NDCG@5, HR@5) are measured on real held-out interactions and do not involve any LLM-based component, providing simulator-independent evidence that the directional feedback induces genuine recommendation improvements rather than artifacts of LLM-based evaluation.

\begin{table*}[t]
\centering
\caption{Overall model performance (A: AlphaEvolve, D: DeepEvolve).}
\resizebox{1.0\linewidth}{!}{%
\begin{tabular}{cc|cccc|c|cccc|cccc|cccc|ccc}
\toprule[1.5pt]
\multirow{2}{*}{Dataset} & \multirow{2}{*}{Metric} & \multicolumn{4}{c|}{NAS} & \multicolumn{4}{c}{Seed Recommender: NCF} & \multicolumn{4}{c}{Seed Recommender: NGCF} & \multicolumn{4}{c}{Seed Recommender: SASRec} & \multicolumn{4}{c}{Seed Recommender: MoRec} \\ 
\cmidrule(lr){3-6} \cmidrule(lr){7-10} \cmidrule(lr){11-14} \cmidrule(lr){15-18} \cmidrule(lr){19-22}
 &  & AutoFIS & NASRec &AutoLossGen& DNS-Rec & NCF & A & D & \textbf{Ours} & NGCF & A & D & \textbf{Ours} & SASRec & A & D & \textbf{Ours} & MoRec & A & D & \textbf{Ours} \\ 
\midrule
\multirow{2}{*}{CDs} & NDCG@5 & 0.2077 & 0.2026 & 0.2123 & 0.2636 & 0.2312 & \uline{0.2556} & 0.2421 & \textbf{0.2723} & 0.3446 & \uline{0.3449} & 0.3280 & \textbf{0.3721} & 0.3559 & 0.3528 & \uline{0.3610} & \textbf{0.3865} & 0.2558 & 0.2414 & \uline{0.3701} & \textbf{0.3977} \\
 & HR@5 & 0.3040 & 0.2989 & 0.3063 & 0.3605 & 0.2979 & 0.3477 & \uline{0.3497} & \textbf{0.3799} & \uline{0.4924} & 0.4896 & 0.4465 & \textbf{0.4974} & 0.4676 & 0.4623 & \uline{0.4870} & \textbf{0.5274} & 0.3779 & 0.3607 & \uline{0.4864} & \textbf{0.5340} \\ 
\midrule
\multirow{2}{*}{Electronics} & NDCG@5 & 0.1753 & 0.1706 & 0.1877 & 0.1598 & 0.1078 & 0.1183 & \uline{0.1817} & \textbf{0.1907} & 0.1531 & \uline{0.1808} & 0.1726 & \textbf{0.1925} & 0.2325 & 0.2063 & \uline{0.2508} & \textbf{0.2600} & 0.1883 & 0.1912 & \uline{0.1938} & \textbf{0.2056} \\
 & HR@5 & 0.2456 & 0.2444 & 0.2512 & 0.2271 &  0.1610 & 0.1733 & \uline{0.2600} & \textbf{0.2714} & 0.2247 & \uline{0.2590} & 0.2385 & \textbf{0.2759} & 0.3208 & 0.2891 & \uline{0.3427} & \textbf{0.3591} & 0.2724 & 0.2675 & \uline{0.2745} & \textbf{0.2921} \\ 
\midrule
\multirow{2}{*}{Office} & NDCG@5 & 0.1714 & 0.1377 &0.1610  & 0.1285 & 0.1620 & \uline{0.1751} & 0.1705 & \textbf{0.1759} & 0.1711 & 0.1799 & \uline{0.1805} & \textbf{0.1930} & 0.1799 & \uline{0.1939} & 0.1816 & \textbf{0.2329} & \uline{0.1851} & 0.1848 & 0.1697 & \textbf{0.1884} \\
 & HR@5 & 0.2490 & 0.2070 & 0.2419 & 0.1879 & 0.2343 & 0.2523 & \uline{0.2548} & \textbf{0.2659} & 0.2494 & \uline{0.2634} & 0.2591 & \textbf{0.2743} & 0.2526 & \uline{0.2750} & 0.2631 & \textbf{0.3218} & \uline{0.2689} & 0.2633 & 0.2447 & \textbf{0.2703} \\ 
\midrule
\multirow{2}{*}{MovieLens} & NDCG@5 & 0.1369 & 0.2916 & 0.2348 & 0.3340 & 0.3413 & \uline{0.3465} & 0.1908 & \textbf{0.3764} & 0.1824 & \uline{0.2355} & 0.2010 & \textbf{0.3588} & 0.5667 & 0.5583 & \uline{0.5722} & \textbf{0.5765} & 0.3796 & 0.4993 & \uline{0.5281} & \textbf{0.5460} \\
 & HR@5 & 0.2099 & 0.4346 & 0.3555 & 0.5046 & 0.4970 & \uline{0.5091} & 0.2735 & \textbf{0.5475} & 0.2876 & \uline{0.3672} & 0.3162 & \textbf{0.5220} & 0.7283 & 0.7199 & \uline{0.7344} & \textbf{0.7366} & 0.5414 & 0.6743 & \uline{0.6967} & \textbf{0.7131} \\ 
\bottomrule[1.5pt]
\end{tabular}%
}
\label{tab:overall_performance}
\vspace{-2ex}
\end{table*}

\smallskip
\noindent\textbf{Implementation Details.}
To ensure fair comparisons, all evolution-based methods are initialized with four distinct seed recommenders: NCF~\cite{he2017neural} (MF-based), NGCF~\cite{wang2019neural} (graph-based), SASRec~\cite{kang2018self} (sequential), and MoRec~\cite{yuan2023go} (multi-modal). We employ GPT-5-mini for planning and retrieval, while utilizing GPT-5 as the coding agent for all frameworks. {We provide additional experiments with different LLMs in App.~\ref{app different llm}.} To account for the computational overhead associated with these LLMs, we provide a comprehensive efficiency analysis, including time cost per iteration and user sampling impact, in App.~\ref{app: eff analysis}. Please refer to the App.~\ref{app: Implementation} and ~\ref{app: seed recommender} for more details regarding the seed recommender and hyper-parameters settings.

\vspace{-1ex}
\subsection{Performance Comparison}
\label{sec: experiment performance comparison}

\noindent\textbf{Overall performance. }
The results of the recommendation task on four datasets are given in Table~\ref{tab:overall_performance}. From the results, we have the following observations: 
\textbf{1)} \proposed~consistently outperforms AlphaEvolve and DeepEvolve across all datasets, regardless of the seed model. These results demonstrate that integrating qualitative critiques with quantitative diagnostics enables effective algorithmic evolution across diverse recommender architectures.
\textbf{2)} AlphaEvolve and DeepEvolve exhibit inconsistency across datasets, with both baselines occasionally underperforming the initial seeds. This inconsistency indicates that without directional feedback, these baselines are limited to trial-and-error searches. Conversely, \proposed~leverages semantic critiques to identify and resolve specific structural deficiencies, leading to more informed and effective improvements.
\textbf{3)} \proposed~significantly surpasses NAS-based baselines, such as AutoFIS and NASRec, in all datasets. This confirms that shifting from a closed, human-defined operator pool to an open-ended program space allows for more expressive logic and structural algorithmic refinement.

\smallskip
\noindent\textbf{User Satisfaction Analysis. }

\begin{wraptable}{r}{0.5\textwidth}
\centering
\vspace{-1ex} 
\captionsetup{font=small}
\caption{Multi-facet satisfaction metrics on SASRec and NCF under two agentic evaluators (Agent4Rec and PUB) (S: Seed Recommender, A: AlphaEvolve, D: DeepEvolve).}
\setlength{\tabcolsep}{4pt}
\resizebox{0.95\linewidth}{!}{
\begin{tabular}{ccccccccccc}
\toprule[1.5pt]
\multirow{2}{*}{\makecell{Seed \\ Recommender}} & \multirow{2}{*}{\makecell{Agentic \\ Evaluator}} & \multirow{2}{*}{Metric} & \multicolumn{4}{c}{CDs} & \multicolumn{4}{c}{Electronics} \\ 
\cmidrule(lr){4-7} \cmidrule(lr){8-11}
 & & & S & A & D & \textbf{Ours} & S & A & D & \textbf{Ours} \\ 
\midrule
\multirow{6}{*}{SASRec} 
 & \multirow{3}{*}{Agent4Rec} 
 & View & 0.372 & 0.378 & \uline{0.379} & \textbf{0.381} & 0.335 & 0.342 & \uline{0.351} & \textbf{0.353} \\
 & & Satisfy & 4.606 & 4.384 & \uline{4.710} & \textbf{5.046} & 4.173 & 4.308 & \uline{4.487} & \textbf{4.502} \\
 & & Depth & 1.926 & 1.830 & \uline{1.952} & \textbf{2.048} & 1.754 & 1.778 & \uline{1.782} & \textbf{1.797} \\ 
\cmidrule(lr){2-11}
 & \multirow{3}{*}{PUB} 
 & View & 0.128 & 0.122 & \uline{0.130} & \textbf{0.136} & 0.133 & 0.134 & \uline{0.141} & \textbf{0.144} \\
 & & Satisfy & 4.630 & 4.506 & \uline{4.810} & \textbf{4.906} & 3.928 & 3.972 & \uline{4.082} & \textbf{4.134} \\
 & & Depth & 1.912 & 1.910 & \uline{1.922} & \textbf{2.018} & 1.856 & 1.888 & \uline{1.904} & \textbf{1.928} \\ 
\midrule
\multirow{6}{*}{NCF} 
 & \multirow{3}{*}{Agent4Rec} 
 & View & 0.365 & 0.354 & \uline{0.369} & \textbf{0.372} & \uline{0.339} & 0.337 & 0.307 & \textbf{0.342} \\
 & & Satisfy & 4.206 & \uline{4.461} & 4.392 & \textbf{4.650} & \uline{4.320} & 4.270 & 4.132 & \textbf{4.354} \\
 & & Depth & 1.776 & \uline{1.882} & 1.840 & \textbf{1.934} & \uline{1.786} & 1.744 & 1.720 & \textbf{1.798} \\ 
\cmidrule(lr){2-11}
 & \multirow{3}{*}{PUB} 
 & View & 0.125 & \uline{0.127} & 0.119 & \textbf{0.134} & \uline{0.135} & 0.125 & 0.124 & \textbf{0.139} \\
 & & Satisfy & 4.488 & \uline{4.650} & 4.422 & \textbf{4.770} & \uline{3.842} & 3.826 & 3.564 & \textbf{3.952} \\
 & & Depth & 1.914 & 1.918 & \uline{1.940} & \textbf{1.952} & \uline{1.874} & 1.768 & 1.754 & \textbf{1.916} \\ 
\bottomrule[1.5pt]
\end{tabular}}
\vspace{-2ex}
\label{tab: user simulation test}
\end{wraptable}
To bridge the gap between static metrics and actual user satisfaction, we adopt the agentic simulation environment from Agent4Rec~\cite{zhang2024generative} and PUB~\cite{ma2025pub} as a scalable proxy for A/B testing.
To mitigate concerns about a simulator-to-simulator evaluation loop, our setup deliberately decouples the optimization-time simulator (SIM) from the evaluation-time simulators along three axes: (i) \textbf{disjoint user populations} -- both Agent4Rec and PUB sample evaluation users that are unseen by SIM during evolution; (ii) \textbf{distinct persona schemes} -- PUB employs Big Five traits, whereas SIM uses Activity/Conformity/Diversity; and (iii) \textbf{independent frameworks} -- both evaluators are external systems not modified by our pipeline. 

In this dynamic environment, generative agents driven by distinct traits (e.g., Big Five) are presented with pages of four items and are asked to decide whether to continue or terminate the session based on relevance and diversity of recommended items.
Following Agent4Rec, we quantify this simulation using three key user-centric metrics: \textbf{View} (item view ratio), \textbf{Satisfy} (a comprehensive score about recommender from 1 to 10), and \textbf{Depth} (the number of pages explored before termination). 
Notably, in the Agent4Rec setup, agents terminate recommendation sessions upon encountering unsatisfactory items, indicating that a higher \textbf{Depth} signifies successful user retention through consistently relevant recommendations.
The results of the user satisfaction are summarized in Table~\ref{tab: user simulation test}. From the results, we have the following observations:
\textbf{1)} \proposed~consistently outperforms all baselines across all satisfaction-oriented metrics (i.e., View, Satisfy, and Depth) under \textbf{both Agent4Rec and PUB}, regardless of the seed model. The gains under PUB are particularly informative: since PUB's Big Five persona scheme is fundamentally distinct from the Activity/Conformity/Diversity scheme used by SIM during evolution, these results confirm that \proposed~does not overfit to the critique style of its own optimization-time simulator. Instead, the directional feedback from $\mathcal{R}_{\text{SIM}}$ enables the models to evolve beyond mere accuracy, substantially improving the perceived quality of recommendations from a user-centric perspective and sustaining user engagement for longer durations.
\textbf{2)} In contrast, AlphaEvolve and DeepEvolve show suboptimal generalization to user-centric metrics and, in some cases, even underperform the initial seed models. This suggests that scalar metric-only optimization is insufficient to capture the complex dynamics of the user experience. Without guidance, the evolved models tend to overfit to narrow numerical targets, which degrades overall user satisfaction.

\subsubsection{\textbf{Codebase Quality Evaluation}} \label{sec: codebase quality}

\begin{wrapfigure}{r}{0.45\textwidth}
  \centering
  \vspace{-3ex}
  \includegraphics[width=0.95\linewidth]{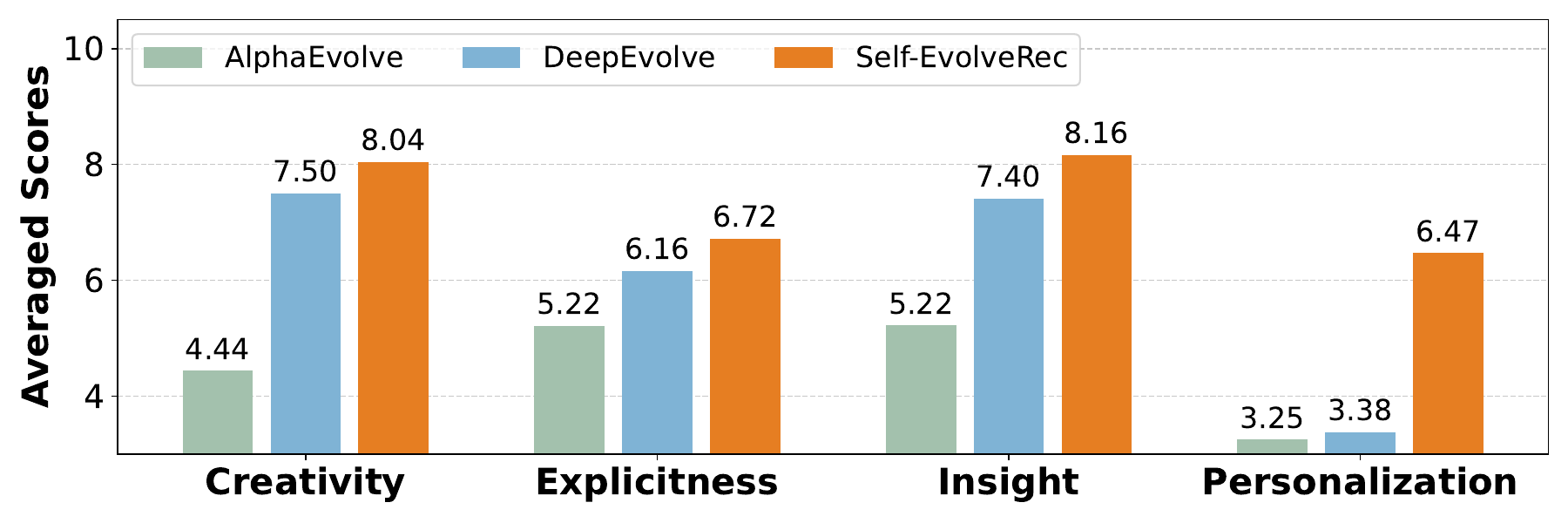}
  \vspace{-1.5ex}
  \captionsetup{font=small}
  \caption
    {LLM-as-a-Judge evaluation of the evolved models.}
    \vspace{-2ex}
    \label{fig: llm_as_judge}
\end{wrapfigure}
To verify substantive algorithmic improvements beyond numerical gains, we employ an LLM-as-a-judge~\cite{zheng2023judging,kim2025beyond,gu2024survey} (averaged scores from GPT-5 and Claude Opus 4.6) to evaluate the recommender code in all evolved codebases of Table~\ref{tab:overall_performance} on a 1-10 scale against the seed codebase, across four dimensions: \textbf{Creativity} (novel mechanisms beyond parameter tuning), \textbf{Explicitness} (interpretability of logic flows), \textbf{Insight} (intent to resolve specific failures, e.g., popularity bias), and \textbf{Personalization} (user-context awareness). Detailed prompts are provided in App.~\ref{app prompts llm-as-a-judge}.
From the results in Figure~\ref{fig: llm_as_judge}, we have the following observations:
\textbf{1)}~\proposed~consistently achieves the highest scores across all criteria. The high scores on Creativity and Insight validate that our directional feedback loop integrating \text{SIM} and \text{DIAG} enables targeted logic refinement, whereas scalar-driven baselines remain limited to trial-and-error without diagnosing underlying issues. \textbf{2)}~\proposed~shows a substantial gain in Personalization (+90\% over baselines), driven by \text{SIM} reflecting specific user needs directly into algorithmic logic, while baselines optimize purely numerical metrics. \textbf{3)} AlphaEvolve underperforms across all criteria due to the absence of external knowledge, underscoring the necessity of retrieval (as in DeepEvolve and~\proposed) for open-ended discovery of novel mechanisms.
\subsection{Ablation Studies}
\vspace{-.5ex}
\begin{wraptable}{r}{0.5\linewidth}
\centering
\vspace{-5ex}
\captionsetup{font=small}
\caption{Ablation studies on the components of~\proposed~(Seed Recommender: SASRec).}
\vspace{-1ex}
\setlength{\tabcolsep}{4pt}
\resizebox{\linewidth}{!}{%
\begin{tabular}{cccccc}
\toprule[1.5pt]
\multirow{2}{*}{Row} & \multirow{2}{*}{Components} & \multicolumn{2}{c}{CDs} & \multicolumn{2}{c}{Electronics} \\
\cmidrule(lr){3-4} \cmidrule(lr){5-6}
 & & NDCG@5 & HR@5 & NDCG@5 & HR@5 \\
\midrule
(1) & {\scriptsize Metric only}            & 0.3610 & 0.4870 & 0.2508 & 0.3427 \\
(2) & {\scriptsize w/ SIM}                 & 0.3751 & 0.5102 & 0.2573 & 0.3551 \\
(3) & {\scriptsize w/ DIAG}                & 0.3676 & 0.4791 & 0.2515 & 0.3449 \\
(4) & {\scriptsize w/ SIM, DIAG}           & 0.3789 & 0.5164 & 0.2584 & 0.3566 \\
(5) & {\scriptsize w/ DIAG, Co-Evolve}     & 0.3727 & 0.5014 & 0.2532 & 0.3520 \\
(6) & {\scriptsize w/ SIM, DIAG, Co-Evolve} & \textbf{0.3865} & \textbf{0.5274} & \textbf{0.2600} & \textbf{0.3591} \\
\bottomrule[1.5pt]
\end{tabular}%
}
\vspace{-2ex}
\label{tab:ablation_study}
\end{wraptable}
To comprehensively evaluate the impact of each component in~\proposed, we conduct ablation studies in Table~\ref{tab:ablation_study}. Note that Row (6) represents the complete framework of~\proposed. We have the following observations: Across all datasets, \textbf{1)} Introducing either the \text{SIM} (Row (2)) or the \text{DIAG} (Row (3)) consistently yields performance gains over the scalar metric only method (Row (1)). Notably, the integration of the \text{SIM} results in a more substantial increase, underscoring its critical role in providing directional feedback by pinpointing specific recommendation failures, which scalar metrics fail to capture. \textbf{2)} Comparing Row (3) and Row (5) demonstrates that static diagnostic criteria limit the potential of code evolution. Since the model architecture continuously shifts, a fixed \text{DIAG} inevitably becomes obsolete. The performance gain in Row (5) implies that dynamically synchronizing the diagnostic logic of \text{DIAG} via co-evolution is essential for maintaining a reliable feedback loop. \textbf{3)} Integrating both the \text{SIM} and the \text{DIAG} (Row (4)), achieves superior performance among all ablated variants, notably outperforming even the co-evolved \text{DIAG} without user feedback (Row (5)). This underscores the importance of combining qualitative user critiques and quantitative diagnostics, which provide recommendation failures and internal structural issues. Furthermore, the complete framework of~\proposed~(Row (6)) achieves the best performance, validating that dynamically aligning the diagnostic logic with both the shifting architecture and the qualitative feedback is necessary to ensure the efficacy of the feedback loop and the evolution of the model.


\subsection{Model Analysis}
\vspace{-0.5ex}
\subsubsection{\textbf{Adaptability to extreme initialization scenarios}}\label{exp: extreme initial}
To further validate the adaptability of~\proposed, we evaluate its performance under two extreme initialization settings: a Random recommender representing a development starting from scratch, and a highly optimized Ensemble (NCF + NGCF + SASRec) reflecting a sophisticated industrial deployment.
We have the following observations in Table~\ref{tab:initial_model}: \textbf{1)} Even starting from a random recommender, \proposed~successfully evolves a fully functional and highly competitive recommender pipeline.
{\proposed~reaches the peak performance at the 8th (CDs) and the 11th (Electronics) iterations, significantly earlier than baselines (13 to 19 iterations).
{Furthermore, in terms of actual evolution time (excluding training) on the random recommender - CDs,~\proposed~efficiently reaches peak performance in 2h 27m compared to DeepEvolve's 3h 2m. While AlphaEvolve is faster (1h 8m), it yields suboptimal performance.}
This confirms that directional feedback systematically constructs valid pipelines by resolving structural deficiencies, avoiding the aimless trial-and-error of scalar-driven methods.}
{Further details regarding the evolution time required to reach peak performance are provided in Appendix~\ref{app: time cost}.}
\textbf{2)} In the ensemble setting, achieving further gains is exceedingly challenging due to the high initial performance and structural complexity of seed recommenders.
We observe that scalar-only baselines, AlphaEvolve and DeepEvolve, struggle to navigate the structural complexity of the ensemble, even degrading the initial performance. In contrast,~\proposed~consistently identifies and resolves latent bottlenecks within the ensemble, demonstrating that directional feedback is effective even in high-performance regimes, where numerical scores alone fail to provide guidance for optimizing such structurally complex systems.

These results highlight the adaptability of \proposed~across diverse industrial scenarios, ranging from the rapid cold-start of new services to the iterative refinement of high-performance models for established domains \cite{cheng2016wide, 10.1145/2843948, wang2022learning, 10.1145/3580305.3599516}.

\begin{table}[]
\centering
\captionsetup{font=small}
\caption{Performance comparison under extreme initialization scenarios (Random and Ensemble). {Peak indicates the iteration number of the best performance.}}
\label{tab:initial_model}
\resizebox{0.875\linewidth}{!}{%
\begin{tabular}{cccccccccccc}
\toprule[1.5pt]
\multirow{2}{*}{Dataset} & \multicolumn{2}{c}{Seed Recommender} & \multicolumn{3}{c}{AlphaEvolve} & \multicolumn{3}{c}{DeepEvolve} & \multicolumn{3}{c}{\textbf{Ours}} \\ 
\cmidrule(lr){2-3} \cmidrule(lr){4-6} \cmidrule(lr){7-9} \cmidrule(lr){10-12}
 & NDCG@5 & HR@5 & NDCG@5 & HR@5 & Peak & NDCG@5 & HR@5 & Peak & NDCG@5 & HR@5 & Peak \\ 
\midrule
\multicolumn{12}{l}{\textit{\textbf{Seed Recommender: Random}}} \\
CDs & 0.0312 & 0.0525 & 0.3430 & 0.4549 & 15 & \uline{0.3766} & \uline{0.4963} & \uline{13} & \textbf{0.3883} & \textbf{0.5165} & \textbf{8}\\
Electronics & 0.0310 & 0.0531 & \uline{0.2037} & \uline{0.2878} & 19 & 0.1972 & 0.2761 & \uline{18} & \textbf{0.2109} & \textbf{0.2952}& \textbf{11} \\ 
\midrule
\multicolumn{12}{l}{\textit{\textbf{Seed Recommender: NCF+NGCF+SASRec}}} \\
CDs & \uline{0.3946} & \uline{0.5179} & 0.3864 & 0.5075 & FAIL & 0.3695 & 0.5002 & \uline{17} & \textbf{0.4105} & \textbf{0.5409} & \textbf{9}\\
Electronics & 0.2385 & 0.3246 & \uline{0.2496} & \uline{0.3386} & \uline{14} & 0.2353 & 0.3240 & FAIL & \textbf{0.2524} & \textbf{0.3426} & \textbf{6}\\ 
\bottomrule[1.5pt]
\end{tabular}%
}
\end{table}


\vspace{-0.75ex}
\subsubsection{\textbf{Removing Feedback-aware Planning \& Retrieval}} To evaluate the contribution of the feedback-aware planning \& retrieval introduced in Sec.~\ref{Sec Evolution Pipeline}, we compare \proposed~against a variant that excludes Planning \& Retrieval (i.e., $\mathcal{R}_\text{Dev}$), denoted as \textbf{w.o. Planning} in Table~\ref{tab:alphaevolve_Ours}.
In this variant, the agent generates code directly based on the directional feedback through the code evolution step as $\mathcal{B}^{(t+1)} = \text{LLM}(\allowbreak \mathcal{I}_{\text{CODE}}, \allowbreak \mathcal{R}_{\text{SIM}}, \allowbreak \mathcal{R}_{\text{DIAG}}, \allowbreak \mathcal{B}_{\text{parent}},\allowbreak \mathcal{P}^{(t)})$ instead of $\mathcal{B}^{(t+1)} = \text{LLM}(\mathcal{I}_{\text{CODE}}, \mathcal{R}_{\text{Dev}}, \mathcal{B}_{\text{parent}}, \mathcal{P}^{(t)})$.
From the results in Table~\ref{tab:alphaevolve_Ours}, we have the following observations: \textbf{1)} w.o. Planning achieves comparable recommendation performance to the original~\proposed.
This implies that the primary driver of performance is the precise identification of failure modes via directional feedback, rather than the incorporation of external research ideas.
\textbf{2)} However, the absence of planning significantly degrades codebase quality.
While Insight remains relatively high (effectively identify what is wrong), w.o. Planning exhibits substantial drops in Creativity and Explicitness.
This suggests that without planning, the agent relies on local heuristic patches rather than systematic, modular designs grounded in external knowledge.
{\textbf{3)} 
While DeepEvolve achieves high Creativity via external knowledge (RAG), w.o. Planning achieves higher Personalization. 
This confirms that the directional feedback from \text{SIM}, which captures specific user needs, is more crucial for personalization than generic knowledge retrievals.}


\begin{table}[t]
\begin{minipage}[t]{0.53\linewidth}
\centering
\captionsetup{font=footnotesize}
\captionof{table}{Performance when Feedback-aware Planning \& Retrieval is removed (Seed Recommender: SASRec). C: Creativity, E: Explicitness, I: Insight, P: Personalization.}
\label{tab:alphaevolve_Ours}
\resizebox{\linewidth}{!}{%
\begin{tabular}{ccccccccc}
\toprule[1.5pt]
\multirow{2}{*}{\textbf{Method}} & \multicolumn{2}{c}{\textbf{CDs}} & \multicolumn{2}{c}{\textbf{Electronics}} & \multicolumn{4}{c}{\textbf{Codebase Quality}} \\ 
\cmidrule(lr){2-3} \cmidrule(lr){4-5} \cmidrule(lr){6-9}
 & NDCG@5 & HR@5 & NDCG@5 & HR@5 & C & E & I & P\\ 
\midrule
AlphaEvolve   & 0.3528 & 0.4623 & 0.2063 & 0.2891 & 4.5 & 5.0 & 5.5 & 3.0\\
DeepEvolve   & 0.3610 & 0.4870 & 0.2508 & 0.3427 & \uline{7.5} & \uline{6.5} & {7.5} & 3.0\\
\proposed   & 0.3865 & \textbf{0.5274} & \textbf{0.2600} & \textbf{0.3591} & \textbf{8.0} & \textbf{7.5} & \textbf{8.5} & \textbf{6.5}\\
\midrule
w.o. Planning & \textbf{0.3988} & \uline{0.5134} & 0.2597 & \uline{0.3568} & 6.5 & {5.0} & \uline{8.0} & \uline{5.0} \\
\bottomrule[1.5pt]
\end{tabular}%
}
\end{minipage}
\hfill
\begin{minipage}[t]{0.45\linewidth}
\centering
\captionsetup{font=footnotesize}
\captionof{table}{Evolving User Simulator. NDCG@5/HR@5 report the performance of evolved NCF, and Reliability denotes the accuracy of identifying the target item from 20 candidates by SIM.}
\label{tab:simulator evolve}
\resizebox{\linewidth}{!}{%
\begin{tabular}{c|ccc||cc}
\toprule
\textbf{Dataset} & \textbf{Strategy} & \textbf{NDCG@5} & \textbf{HR@5} & \textbf{SIM Type} & \textbf{Reliability} \\
\midrule
\multirow{3}{*}{\textbf{CDs}} 
& -                 & -      & -      & NCF (Base.)  & 0.2882 \\
& Fixed SIM         & 0.2723 & \textbf{0.3799} & Initial SIM  & 0.3473 \\
& Evolved SIM       & \textbf{0.2745} & 0.3791 & Evolved SIM  & \textbf{0.3610} \\
\midrule
\multirow{3}{*}{\textbf{Elec.}} 
& -                 & -      & -      & NCF (Base.)  & 0.2191 \\
& Fixed SIM         & \textbf{0.1907} & 0.2714 & Initial SIM  & 0.2238 \\
& Evolved SIM       & 0.1891 & \textbf{0.2744} & Evolved SIM  & \textbf{0.2341} \\
\bottomrule
\end{tabular}%
}
\end{minipage}
\vspace{-1ex}
\end{table}

\vspace{-0.5ex}
\subsubsection{\textbf{
Trustworthiness of the User Simulator}}
Since SIM's qualitative critiques drive the entire directional feedback loop, any factual hallucination---such as fabricated user preferences or interactions absent from $\mathcal{H}_u$---would propagate as misleading guidance throughout evolution. We therefore examine whether SIM's feedback is grounded in actual user behavior. Following GISTBench~\cite{fostiropoulos2026gistbench}, we measure \textbf{Interest Groundedness} (precision) on all simulation feedback generated along the trajectories in Table~\ref{tab:overall_performance}, achieving 0.891 (CDs), 0.843 (Electronics), and 0.851 (Office). These results confirm that SIM rarely fabricates preferences; its critiques are anchored in observable user history, ensuring that the directional feedback driving~\proposed~reflects realistic user intent rather than LLM-induced artifacts. A full IG precision/recall analysis, together with a complementary discussion of how DIAG calibrates SIM's occasional semantic over-generalization (e.g., flagging same-artist exposure as redundancy), is provided in App.~\ref{app: hallucination groundedness}.


\vspace{-0.5ex}
\subsubsection{\textbf{
Evolving User Simulator}}\label{exp sim evolve}
Although our framework employs a fixed user simulator for efficiency, we further investigate the impact of evolving the user simulator alongside the codebase.
Analogous to Sec.~\ref{Sec Co-evolution}, the co-evolution of the simulator follows the same structured workflow of analysis, retrieval, and planning.
Specifically, the LLM analyzes the current simulator and synthesizes a development plan $\mathcal{R}_{\text{Dev-SIM}}$ using retrieved methodologies $\mathcal{K}_{\text{SIM}}$, yielding the evolved simulator: $\text{SIM}^{(t+1)} = \text{LLM}(\mathcal{I}_{\text{Code-SIM}},\allowbreak \mathcal{R}_{\text{Dev-SIM}}, \allowbreak \text{SIM}_{\text{parent}}, \allowbreak \mathcal{P}^{(t)})$.
In Table~\ref{tab:simulator evolve}, we evaluate simulator reliability by measuring the accuracy of identifying the target item from 20 candidates, utilizing a simple recommender model, NCF, as a baseline.
While the recommendation performance remains comparable, the \text{SIM}'s accuracy reveals the following observation: \textbf{1)} The comparable recommendation performance indicates that the initial \text{SIM} is already effective for modeling complex user preferences, achieving the Reliability superior to the NCF. This result indicates that the \text{SIM} provides reliable, high-quality directional feedback that captures complex user intent in decision making processes on given recommendation list. Consequently, even without evolving the simulator, the feedback signals from $\mathcal{R}_{\text{SIM}}$ are already robust enough to guide the~\proposed~toward an optimal codebase. \textbf{2)} Nevertheless, evolution on \text{SIM} further elevates the simulator's reliability. The evolved \text{SIM} achieves an even higher Reliability compared to its initial state. This confirms that while the resulting recommendation scores are similar on the evolved codebase, the evolutionary process constructs a statistically more trustworthy feedback, ensuring that the directional feedback is grounded in realistic user behavior patterns.

\subsubsection{\textbf{Case Study: Validating Reliability of Co-evolved Diagnostic Tool}}
\begin{wrapfigure}{r}{0.4\textwidth}
\vspace{-2ex}
\centering
\includegraphics[width=0.99\linewidth]{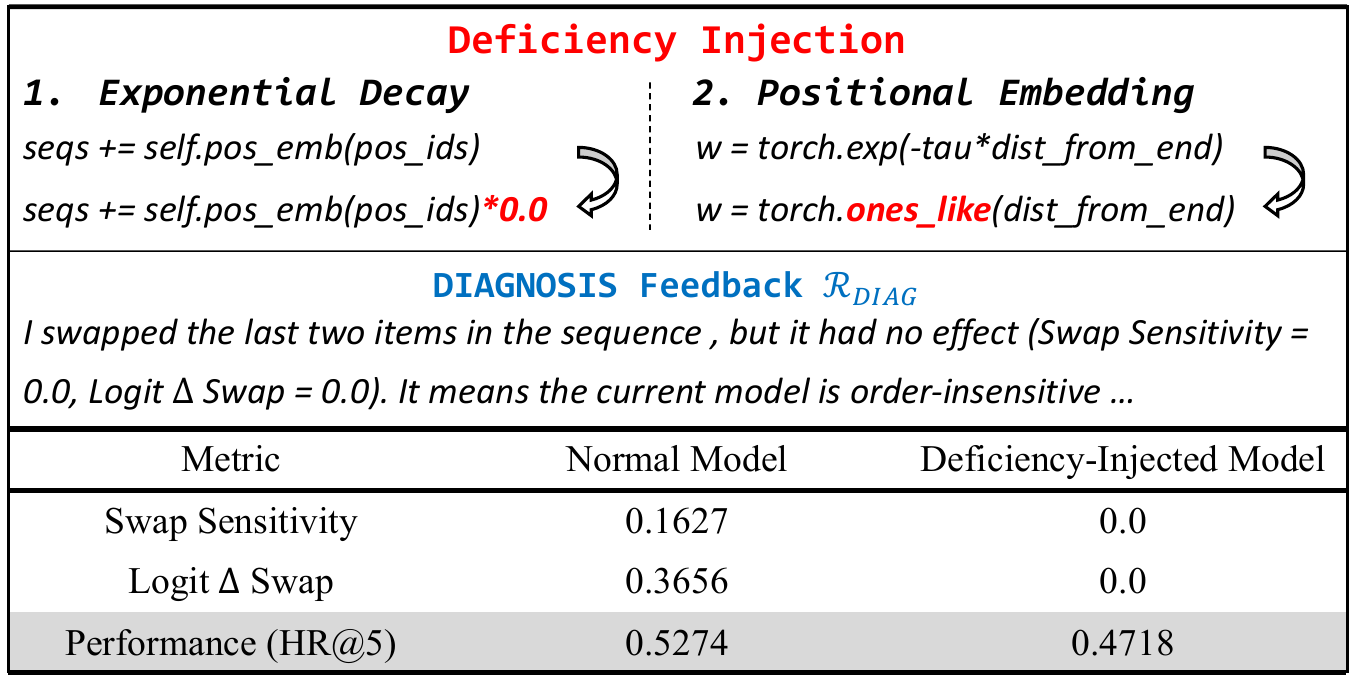}
    \vspace{-1ex}
    \captionsetup{font=small}
    \caption
    {Case study on Diagnosis Tool - Model Co-Evolution on CDs dataset (Seed Recommender: SASRec).}
    \vspace{-2.5ex}
    \label{fig: diagnosis_casestudy}
\end{wrapfigure}
To verify the functional reliability and interpretability of the co-evolved \text{DIAG}, we conduct a case study through an {induced deficiency experiment}. 
Specifically, we evaluate whether the autonomously generated diagnostic metrics can accurately pinpoint logic-level deficiencies intentionally injected in the evolved models.
As shown in Figure~\ref{fig: diagnosis_casestudy}, we injected deficiencies into an evolved SASRec to force order-insensitivity by: (i) removing positional embeddings, (ii) bypassing the exponential decay module.
The co-evolved \text{DIAG} successfully identifies these deficiencies via newly generated metrics, \textbf{Swap Sensitivity} (measuring ranking shifts after swapping last two items in the interaction sequence) and \textbf{Logit $\Delta$ Swap} (quantifies the deviation in predicted logits for the target item after the swap), which were not present in the initial seed \text{DIAG}. While the normal model exhibits high sensitivity to item order, the deficiency-injected model shows near-zero sensitivity, leading co-evolved \text{DIAG} correctly diagnosed as "Order-insensitive" in $\mathcal{R}_{\text{DIAG}}$. Notably, this diagnostic signal directly correlates with the sharp performance decline, demonstrating that co-evolved \text{DIAG} can explain the root causes of low performance through structural verification.
These results demonstrate the capability to detect complex failure modes, highlighting the pivotal role of Diagnosis Tool - Model Co-Evolution in maintaining a grounded feedback loop.
Additional case studies on MoRec and evolutionary trajectories are provided in App.~\ref{App: Case Study Injection} and App.~\ref{app: case study trajectory}, respectively.

\section{Acknowledgements}
This work was supported by National IT Industry Promotion Agency(NIPA) grant funded by the Korea government(MSIT) (AI-Based Dark Factory Design and Integration Solution, No. RS-2026-25621604), the National Research Foundation of Korea(NRF) grant funded by the Korea government(MSIT) (RS-2026-25607361), and National Research Foundation of Korea(NRF) funded by Ministry of Science and ICT (RS-2022-NR068758).

\vspace{-1ex}
\section{Conclusion}
\label{sec conclusion}
\vspace{-0.5ex}

In this paper, we propose~\proposed, a novel LLM-driven code evolution framework for automating the design of recommender systems. The main idea is to overcome the limitations of existing scalar metric-based code optimization by establishing a directional feedback loop that integrates qualitative critiques from a User Simulator with quantitative verification from a Model Diagnosis Tool. To ensure that this feedback loop remains valid as the recommendation pipeline evolves, we further introduce a Diagnosis Tool - Model Co-Evolution strategy that dynamically adapts diagnostic criteria to structural shifts in the codebase. Extensive experiments demonstrate that~\proposed~significantly outperforms state-of-the-art NAS and LLM-driven evolutionary baselines in both standard recommendation accuracy and multi-faceted user satisfaction, while also exhibiting robust adaptability across diverse initialization settings -- from scratch scenarios to complex ensembles -- relevant to practical industrial deployment. 
While our agentic evaluation deliberately decouples optimization-time and evaluation-time simulators, we acknowledge that fully validating user-centric improvements ultimately requires real-world A/B testing, which we leave for future deployment studies. 


\clearpage
\bibliographystyle{neurips_2024}
\bibliography{sample-base}


\newpage
\section*{NeurIPS Paper Checklist}

The checklist is designed to encourage best practices for responsible machine learning research, addressing issues of reproducibility, transparency, research ethics, and societal impact. Do not remove the checklist: {\bf The papers not including the checklist will be desk rejected.} The checklist should follow the references and follow the (optional) supplemental material.  The checklist does NOT count towards the page
limit. 

Please read the checklist guidelines carefully for information on how to answer these questions. For each question in the checklist:
\begin{itemize}
    \item You should answer \answerYes{}, \answerNo{}, or \answerNA{}.
    \item \answerNA{} means either that the question is Not Applicable for that particular paper or the relevant information is Not Available.
    \item Please provide a short (1--2 sentence) justification right after your answer (even for \answerNA). 
\end{itemize}

{\bf The checklist answers are an integral part of your paper submission.} They are visible to the reviewers, area chairs, senior area chairs, and ethics reviewers. You will also be asked to include it (after eventual revisions) with the final version of your paper, and its final version will be published with the paper.

The reviewers of your paper will be asked to use the checklist as one of the factors in their evaluation. While \answerYes{} is generally preferable to \answerNo{}, it is perfectly acceptable to answer \answerNo{} provided a proper justification is given (e.g., error bars are not reported because it would be too computationally expensive'' or ``we were unable to find the license for the dataset we used''). In general, answering \answerNo{} or \answerNA{} is not grounds for rejection. While the questions are phrased in a binary way, we acknowledge that the true answer is often more nuanced, so please just use your best judgment and write a justification to elaborate. All supporting evidence can appear either in the main paper or the supplemental material, provided in appendix. If you answer \answerYes{} to a question, in the justification please point to the section(s) where related material for the question can be found.

IMPORTANT, please:
\begin{itemize}
    \item {\bf Delete this instruction block, but keep the section heading ``NeurIPS Paper Checklist"},
    \item  {\bf Keep the checklist subsection headings, questions/answers and guidelines below.}
    \item {\bf Do not modify the questions and only use the provided macros for your answers}.
\end{itemize}


\begin{enumerate}

\item {\bf Claims}
    \item[] Question: Do the main claims made in the abstract and introduction accurately reflect the paper's contributions and scope?
    \item[] Answer: \answerYes{} 
    \item[] Justification: Yes, we have clearly described the paper's contributions and scope in the abstract and introduction.

\item {\bf Limitations}
    \item[] Question: Does the paper discuss the limitations of the work performed by the authors?
    \item[] Answer: \answerYes{} 
    \item[] Justification: We have described future work related to paper's limitations in the Sec.~\ref{sec conclusion}.

\item {\bf Theory assumptions and proofs}
    \item[] Question: For each theoretical result, does the paper provide the full set of assumptions and a complete (and correct) proof?
    \item[] Answer: \answerNA{} 
    \item[] Justification: The paper does not include theoretical assumptions.

    \item {\bf Experimental result reproducibility}
    \item[] Question: Does the paper fully disclose all the information needed to reproduce the main experimental results of the paper to the extent that it affects the main claims and/or conclusions of the paper (regardless of whether the code and data are provided or not)?
    \item[] Answer: \answerYes{} 
    \item[] Justification: We include the source code for our proposed work via an external link, along with the implementation details. Additionally, we include the prompts for LLMs in App.~\ref{app prompts}.

\item {\bf Open access to data and code}
    \item[] Question: Does the paper provide open access to the data and code, with sufficient instructions to faithfully reproduce the main experimental results, as described in supplemental material?
    \item[] Answer: \answerYes{} 
    \item[] Justification: All datasets are open-access, and we have provided the pre-processed data along with the source code.

\item {\bf Experimental setting/details}
    \item[] Question: Does the paper specify all the training and test details (e.g., data splits, hyperparameters, how they were chosen, type of optimizer) necessary to understand the results?
    \item[] Answer: \answerYes{} 
    \item[] Justification: We have described the evaluation protocol and implementation details in Sec.~\ref{sec: exp}.

\item {\bf Experiment statistical significance}
    \item[] Question: Does the paper report error bars suitably and correctly defined or other appropriate information about the statistical significance of the experiments?
    \item[] Answer: \answerNo{} 
    \item[] Justification: Due to the substantial API costs associated with the iterative model evolution process, we do not conduct repeated experimental runs. Instead, we demonstrate the robustness of our claims by evaluating our method and baselines across various seed recommenders and multiple datasets. Furthermore, experiments involving different LLM backbones are detailed in App.~\ref{app different llm}.

\item {\bf Experiments compute resources}
    \item[] Question: For each experiment, does the paper provide sufficient information on the computer resources (type of compute workers, memory, time of execution) needed to reproduce the experiments?
    \item[] Answer: \answerYes{} 
    \item[] Justification: We include details regarding GPU and LLM resources in the Implementation Details (Sec.~\ref{sec: exp} and App.~\ref{app: Implementation}).
    
\item {\bf Code of ethics}
    \item[] Question: Does the research conducted in the paper conform, in every respect, with the NeurIPS Code of Ethics \url{https://neurips.cc/public/EthicsGuidelines}?
    \item[] Answer: \answerYes{} 
    \item[] Justification: Our work strictly follows the NeurIPS Code of Ethics.

\item {\bf Broader impacts}
    \item[] Question: Does the paper discuss both potential positive societal impacts and negative societal impacts of the work performed?
    \item[] Answer: \answerNA{} 
    \item[] Justification: As our work focuses on the automatic refinement of recommender pipelines, it does not raise any societal impact concerns.
    
\item {\bf Safeguards}
    \item[] Question: Does the paper describe safeguards that have been put in place for responsible release of data or models that have a high risk for misuse (e.g., pre-trained language models, image generators, or scraped datasets)?
    \item[] Answer: \answerNA{} 
    \item[] Justification: This paper does not pose any risks related to this issue.

\item {\bf Licenses for existing assets}
    \item[] Question: Are the creators or original owners of assets (e.g., code, data, models), used in the paper, properly credited and are the license and terms of use explicitly mentioned and properly respected?
    \item[] Answer: \answerYes{} 
    \item[] Justification: We have properly cited all relevant literature and data sources.

\item {\bf New assets}
    \item[] Question: Are new assets introduced in the paper well documented and is the documentation provided alongside the assets?
    \item[] Answer: \answerYes{} 
    \item[] Justification: We provide detailed and well-documented documented source code via an external repository link, alongside full implementation details in App.~\ref{app: Implementation}.

\item {\bf Crowdsourcing and research with human subjects}
    \item[] Question: For crowdsourcing experiments and research with human subjects, does the paper include the full text of instructions given to participants and screenshots, if applicable, as well as details about compensation (if any)? 
    \item[] Answer: \answerNA{} 
    \item[] Justification: This research does not involve human participants.

\item {\bf Institutional review board (IRB) approvals or equivalent for research with human subjects}
    \item[] Question: Does the paper describe potential risks incurred by study participants, whether such risks were disclosed to the subjects, and whether Institutional Review Board (IRB) approvals (or an equivalent approval/review based on the requirements of your country or institution) were obtained?
    \item[] Answer: \answerNA{} 
    \item[] Justification: This research does not involve human participants.

\item {\bf Declaration of LLM usage}
    \item[] Question: Does the paper describe the usage of LLMs if it is an important, original, or non-standard component of the core methods in this research? Note that if the LLM is used only for writing, editing, or formatting purposes and does \emph{not} impact the core methodology, scientific rigor, or originality of the research, declaration is not required.
    \item[] Answer: \answerYes{} 
    \item[] Justification: Our work relies on LLMs as core components for LLM-driven code evolution. We provide the detailed prompts utilized in App.~\ref{app prompts}. Furthermore, we validated our method across different LLMs in App.~\ref{app different llm}, and conducted experiments in App.~\ref{app: hallucination groundedness} to investigate the hallucination issues arising from LLM dependency.

\end{enumerate}

\clearpage
\appendix

\section{Related Work}
\label{app related work}
\noindent\textbf{Neural Architecture Search for Recommender Systems. }
Neural Architecture Search (NAS) has been increasingly explored in recommender systems to automate the design of feature interactions and model architectures.
Early studies like AutoFIS~\cite{liu2020autofis} replace discrete interaction feature choices with learnable gating to filter redundant interactions.
Later methods extend NAS to backbone design. AutoCTR~\cite{song2020towards} uses evolutionary search to assemble operator blocks into a DAG, whereas DNAS~\cite{krishna2021differentiable}, NASRec~\cite{zhang2023nasrec} and DNS-Rec~\cite{zhang2024dns} leverage weight-sharing supernets for scalability.
Automation has also expanded to non-architectural components. 
AutoLossGen~\cite{li2022autolossgen} casts loss function design as an automated search problem, using reinforcement learning (RL) to explore loss formulations composed of basic mathematical operators (e.g., addition, log, multiplication).
Nevertheless, these approaches remain bounded by pre-defined operator sets and wiring rules, restricting innovation to selection/parameterization within a fixed search space rather than open-ended synthesis of new procedural algorithms.

\smallskip
\noindent\textbf{LLM-driven Code Evolution. }
To overcome the limitations of fixed search spaces, recent work has explored LLM-driven code evolution, which shifts the optimization target from parameters to open-ended programs.
FunSearch~\cite{romera2024mathematical} utilizes an LLM-evaluator loop to discover interpretable algorithms for mathematical tasks, while Eureka~\cite{ma2023eureka} automates RL reward engineering by iteratively refining code based on execution feedback.
Building on this line, AlphaEvolve~\cite{novikov2025alphaevolve} generalizes the evolutionary loop into an autonomous coding pipeline that iteratively edits and tests code using evaluator feedback, scaling to more complex algorithmic optimization tasks.
More recently, DeepEvolve~\cite{liu2025scientific} integrates retrieval-augmented generation, leveraging external knowledge to systematically inform hypothesis generation and implementation for scientific discovery tasks.
However, most existing approaches are guided primarily by a single scalar metric (e.g., accuracy or success rate), which provides limited diagnostic insight into user-centric failure modes such as bias, off-topic, or lack of diversity, which are key considerations for holistic recommender-system optimization.

\smallskip
\noindent\textbf{LLM-based User Simulation. }
LLM-based user simulation has emerged as a promising alternative to static recommendation metrics and costly online A/B testing.
Initial approaches focused on realistic persona construction. Agent4Rec~\cite{zhang2024generative} models social traits like conformity from real-world data, while Profile-aware simulators~\cite{fabbri2025evaluating} utilize natural language summaries of user history to align with human judgment. Enhancing psychological fidelity, PUB~\cite{ma2025pub} further integrates Big Five personality traits~\cite{Goldberg1992THEDO, Roccas2002TheBF} to replicate diverse interaction patterns. Recent research shifts focus to utilizing user simulators for system optimization, and SimUSER~\cite{bougie2025simuser} identifies self-consistent personas with specialized perception and memory modules to serve as believable human proxies. Meanwhile, RecoWorld~\cite{liu2025recoworld} establishes a proactive feedback loop, where the simulator explicitly signals user states (e.g., boredom) to guide the recommender's adaptation.

\section{Traits}
\label{app persona}
\noindent\textbf{Activity (Engagement Level).}
Activity quantifies the degree of a user’s engagement with the recommender system. Since user-item interactions are typically sparse, we define activity as the cardinality of the user’s interaction history:
\begin{equation}
T_{\text{act}}(u) = |\mathcal{H}_u|.
\end{equation}
Users with lower $T_{\text{act}}(u)$ values correspond to passive users who interact infrequently with recommended items, whereas higher values indicate highly engaged users with rich interaction histories.

\noindent\textbf{Conformity (Mainstream Adherence).}
Conformity measures the extent to which a user’s preferences align with global public consensus. This trait captures whether a user follows mainstream tastes or exhibits individualized preferences. We define conformity as the mean squared deviation between the user’s rating $r_{u,v}$ and the global average rating $\bar{r}_v$ of item $v$:
\begin{equation}
T_{\text{conf}}(u) = \frac{1}{|\mathcal{H}_u|}
\sum_{v \in \mathcal{H}_u} (r_{u,v} - \bar{r}_v)^2,
\end{equation}
where $\bar{r}_v$ denotes the average rating of item $v$ across all users. A lower conformity value implies that the user’s preferences closely align with popular sentiment, while a higher value reflects more distinctive and personalized tastes.

\noindent\textbf{Diversity (Interest Breadth).}
Diversity characterizes the breadth of a user’s interests across item categories. We define this trait as the number of unique categories associated with the items in the user’s interaction history:
\begin{equation}
T_{\text{div}}(u) = \big| \{ c_v \mid v \in \mathcal{H}_u \} \big|.
\end{equation}
Users with lower $T_{\text{div}}(u)$ values tend to focus on a narrow set of categories, whereas higher values indicate a preference for exploring a broader and more diverse range of categories.

We categorized each user trait into three distinct levels, defined as follows:
\begin{itemize}[leftmargin=0.25cm]
    \item \textbf{Activity}: 
    \begin{itemize}
        \item HIGH: Frequently interacts with the system and maintains a high volume of engagement with recommendations.
        \item MID: Interacts moderately, primarily when items strictly align with personal preferences.
        \item LOW: Rarely interacts with the system and does not interact if recommendations are not relevant to their interests.
    \end{itemize}
    \item \textbf{Conformity}: 
    \begin{itemize}
        \item HIGH: Heavily influenced by popularity and public ratings; tends to follow mainstream trends.
        \item MID: Considers both popularity and personal taste, balancing trends with individual preferences.
        \item LOW: Ignores popularity and trends, evaluating items purely based on intrinsic personal preference.
    \end{itemize}
    \item \textbf{Diversity}: 
    \begin{itemize}
        \item HIGH: Seeks high variety and novelty, enjoying the exploration of diverse categories and new styles.
        \item MID: Mostly consumes preferred categories but occasionally explores similar alternatives.
        \item LOW: Sticks strictly to a narrow set of familiar categories and avoids exploration.
    \end{itemize}
\end{itemize}

\section{Implementation Details}
\label{app: Implementation}
Regarding the baseline implementation of AlphaEvolve, due to the unavailability of the official code, we utilized OpenEvolve~\cite{openevolve}, an open-source implementation, following prior work~\cite{liu2025scientific}. In our evolutionary framework, we set the maximum evolution steps to 21 across all LLM-driven evolutionary frameworks. For all LLM-driven evolutionary frameworks, we set the maximum evolution iterations to 21. For NAS baselines, we configured the search epochs to 5 for AutoFIS and 1 for NASRec, following the hyper-parameter setting in NASRec~\cite{zhang2023nasrec}. We employ GPT-5-mini for the User Simulator (\text{SIM}) and set the number of sampled users to $|\mathcal{U}_{\text{sample}}|=20$ (refer to App.~\ref{app: user sample} for an analysis of the number of sampled users).
To ensure a fair comparison, we uniformly configured all recommender models—including the retraining phase of NAS models—with a user/item embedding dimension of 50, a batch size of 128, and a learning rate of 0.001. The maximum number of epochs was set to 300 for both standard training and the NAS retraining stage. All experiments were conducted on a single NVIDIA GeForce A6000 (48GB) GPU.

\begin{figure}[t]
    \centering
    \begin{minipage}{0.49\linewidth}
        \centering
        \includegraphics[width=\linewidth]{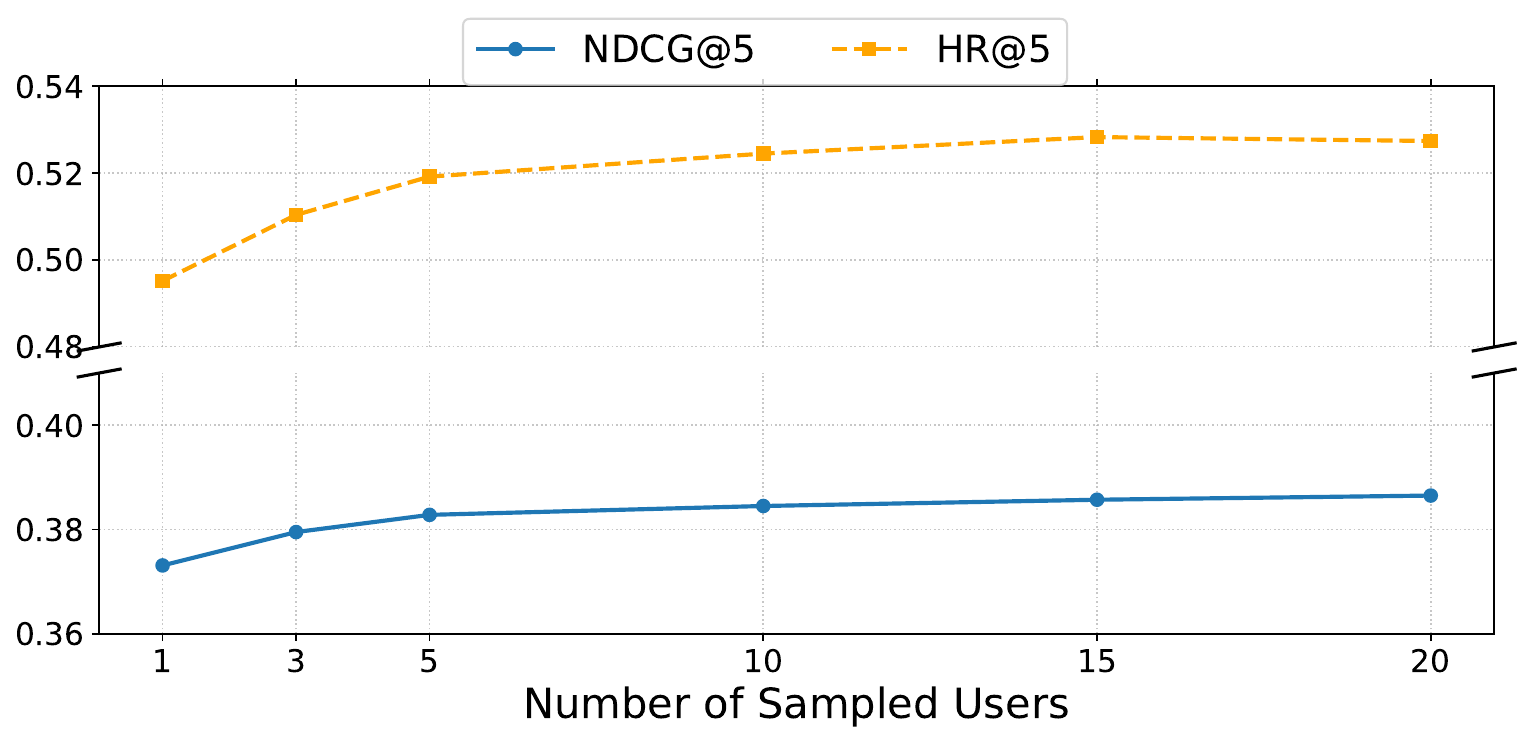}
        \caption{Recommendation performance over number of sampled user $\mathcal{U}_{\text{sample}}$ on CDs dataset (Seed Recommender: SASRec).}
        \label{fig:sample_size}
    \end{minipage}
    \hfill
    \begin{minipage}{0.49\linewidth}
        \centering
        \includegraphics[width=\linewidth]{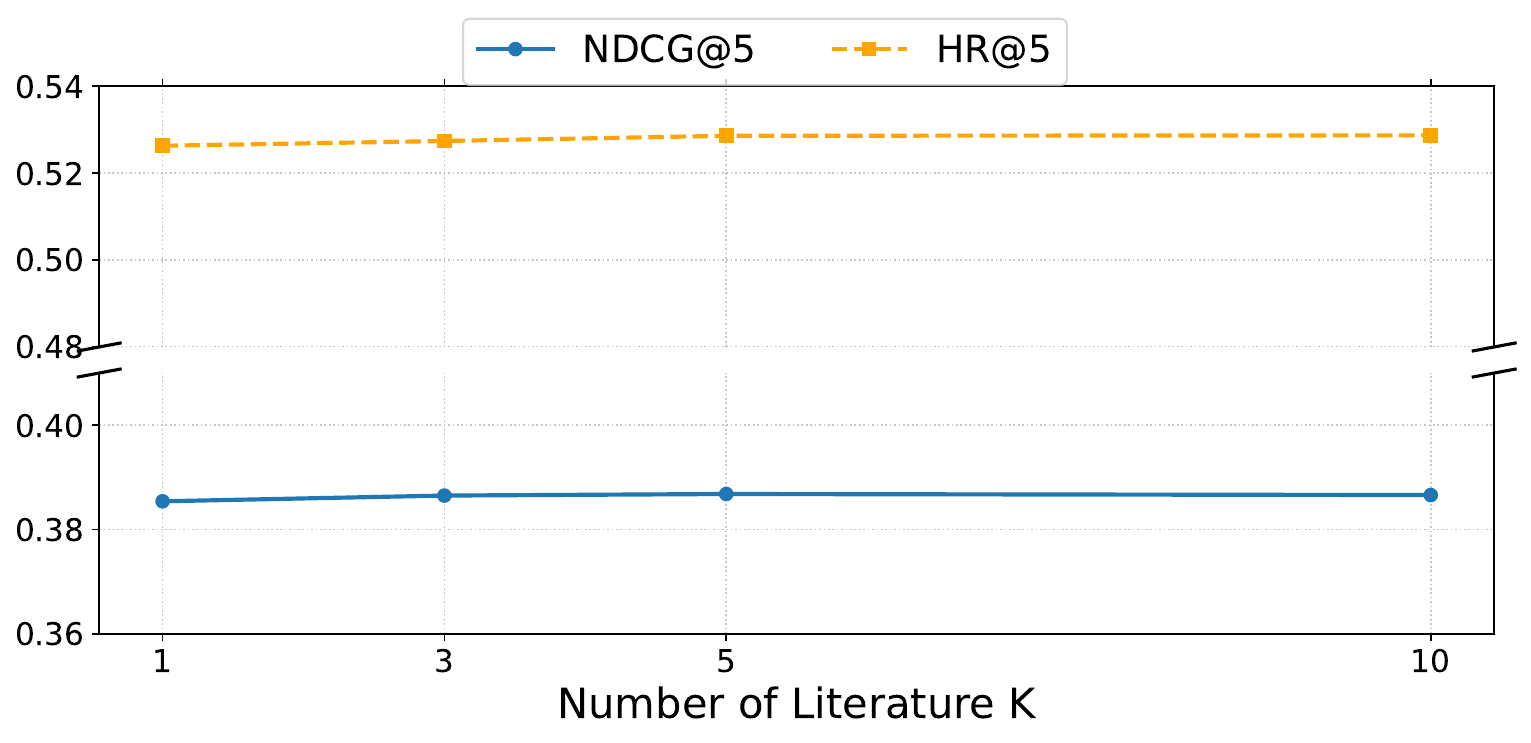}
        \caption{Recommendation performance over number of literature $\mathcal{K}$ on CDs dataset (Seed Recommender: SASRec).}
        \label{fig:literature_sample}
    \end{minipage}
\end{figure}

\section{Datasets}
\label{app dataset}
Table~\ref{app tab dataset} shows the statistics of the dataset after preprocessing.

\begin{table}[t]
\centering
\caption{Statistics of datasets after preprocessing.}
\label{app tab dataset}
\resizebox{0.6\linewidth}{!}{%
\begin{tabular}{ccccc}
\toprule[1.5pt]
Dataset & CDs & Electronics & Office & MovieLens \\
\midrule
\# Users & 14,335 & 32,232 & 20,147 & 6,040 \\ 
\midrule
\# Items & 11,436 & 17,695 & 10,470 & 3,952 \\ \midrule
\# Interactions  & 126,225 & 257,850 & 145,927 & 1,000,209\\
\bottomrule[1.5pt]
\end{tabular}%
}
\end{table}

\section{Baselines and Seed Recommender}
\subsection{Baselines}
\label{app baseline}
\noindent(1) Neural Architecture Search Baselines
\begin{itemize}[leftmargin=0.5cm]
\item AutoFIS~\cite{liu2020autofis} automatically identifies essential feature interactions by employing learnable gates to prune redundant feature combinations.
\item NASRec~\cite{zhang2023nasrec} leverages a weight-sharing supernet to efficiently search for optimal full architectures.
\item AutoLossGen~\cite{li2022autolossgen} automatically searches for effective loss functions tailored to different recommendation models and datasets.
\item DNS-Rec~\cite{zhang2024dns} searches compact architectures for sequential recommender systems under data-aware and resource-aware constraints.
\end{itemize}
\noindent(2) LLM-driven Code Evolution Baselines
\begin{itemize}[leftmargin=0.5cm]
\item AlphaEvolve~\cite{novikov2025alphaevolve} orchestrates an autonomous LLM-driven evolutionary pipeline to iteratively evolve and optimize algorithmic codebases.
\item DeepEvolve~\cite{liu2025scientific} integrates retrieval-augmented generation (RAG) into the evolutionary loop to guide code optimization with external scientific knowledge.
\end{itemize}
\subsection{Seed Recommender}
\label{app: seed recommender}
\begin{itemize}[leftmargin=0.5cm]
\item NCF~\cite{he2017neural} is a pioneering neural collaborative filtering framework which combines multi-layer perceptron (MLP) to learn user-item interactions.
\item NGCF~\cite{wang2019neural} is a graph-based model that explicitly encodes collaborative signals in the embedding space by propagating embeddings on the user-item bipartite graph
\item SASRec~\cite{kang2018self} is a sequential recommendation model leveraging self-attention mechanisms to dynamically capture user interests from interaction sequences.
\item MoRec~\cite{yuan2023go} is a multi-modal framework that utilizes pre-trained encoders (e.g., SBERT) to initialize item embeddings with textual features. Following prior work~\cite{10.1145/3637528.3671931}, we employ SASRec as the backbone architecture for MoRec.
\end{itemize}

\section{Detailed Implementation of the Efficient Variant}
\label{app: zero cost detail}
In this appendix, we provide the precise formulation of the zero-cost signals introduced in Sec.~\ref{Sec efficient variant}, along with the corresponding adaptations of the Diagnosis Tool (DIAG) and the User Simulator (SIM).

\subsection{Detailed Formulation of Zero-Cost Signals.} 
Let $U \in \mathbb{R}^{B \times d}$ and $P \in \mathbb{R}^{B \times d}$ denote the user and positive item representation matrices from a single forward pass on a mini-batch, where $B$ is the batch size and $d$ is the embedding dimension. Let $I = U \odot P \in \mathbb{R}^{B \times d}$ be the interaction representation, and let $\{\sigma_i(M)\}$ denote the singular values of a matrix $M$ in decreasing order. \textbf{1)} The effective ranks $ie$ and $ue$~\citep{roy2007effective} are defined as the exponentiated Shannon entropy of the normalized singular value distribution:
\begin{equation}
\text{erank}(M) = \exp\left(-\sum_i p_i(M) \log p_i(M)\right), \quad p_i(M) = \frac{\sigma_i(M)}{\sum_j \sigma_j(M)},
\end{equation}
with $ie = \text{erank}(I)/min(B,d)$ and $ue = \text{erank}(U)/min(B,d)$. Both take values in $[0,1]$, and a higher value indicates that the representation occupies more independent dimensions. \textbf{2)} The top-1 singular value share $sv = p_1(I) = \sigma_1(I) / \sum_j \sigma_j(I)$ measures the fraction of total spectral energy concentrated in the largest singular value of $I$, directly diagnosing rank collapse~\citep{dong2021attention}.
\textbf{3)} The mean parameter gradient norm $g$ is computed by performing a loss function on the mini-batch, backpropagating to obtain $\nabla_\theta \mathcal{L}$, and averaging the per-parameter gradient norms. This serves as a proxy for the model's initial trainability and information flow~\citep{abdelfattah2021zero}. Finally, we design the proxy $\mathcal{M}_{\text{zero}} = ie + ue - sv - |\log g +3|$ as a surrogate for the standard recommendation metric (e.g., NDCG).

\subsection{Adaptation of the Diagnosis Tool (DIAG).} 
The standard DIAG (Sec.~\ref{Sec Diagnosis Tool}) relies on post-training probes such as embedding collapse, which require fully trained model parameters and therefore cannot be computed under the zero-cost regime. To resolve this, the seed diagnosis tool $\text{DIAG}^{(0)}$ in the efficient variant directly leverages the individual sub-signals of $\mathcal{M}_{\text{zero}}$ as foundational probes for assessing structural health: $ie$ and $ue$ serve as proxies for representation diversity, $sv$ for rank collapse detection, and $g$ for trainability assessment. As the codebase evolves, the Diagnosis Tool - Model Co-Evolution mechanism (Sec.~\ref{Sec Co-evolution}) continues to update these probes accordingly.

\subsection{Adaptation of the User Simulator (SIM).} 
The original SIM (Sec.~\ref{Sec User Simulator}) requires explicit recommendation lists $\mathcal{A}^u$ from a fully trained model to evaluate user-perceived quality. Since $\mathcal{A}^u$ is unavailable without full training, we adapt SIM via a \textit{code-to-intent} summarization strategy, where an LLM summarizes the current codebase $\mathcal{B}^{(t)}$ into a high-level architectural intent. This intent is provided to SIM together with the zero-cost signals, allowing SIM to assess whether the proposed structural changes align with the user's persona $\mathcal{T}_u$ and history $\mathcal{H}_u$, and to generate qualitative feedback $\mathcal{R}_{\text{SIM}}$ (e.g., \textit{``Please capture full multi-category item signals more.''}).

\section{Additional Experiments}

\subsection{{Adaptability to Different LLMs}}
\label{app different llm}
\begin{wraptable}{r}{0.5\linewidth}
\vspace{-8ex}
\centering
\captionsetup{font=small}
\caption{Performance comparison of different LLMs.}
\setlength{\tabcolsep}{4pt}
\resizebox{1.0\linewidth}{!}{
\begin{tabular}{ccccccccc}
\toprule[1.5pt]
\multirow{2}{*}{Seed Model} & \multirow{2}{*}{LLMs} & \multirow{2}{*}{Metric} & \multicolumn{3}{c}{CDs} & \multicolumn{3}{c}{Electronics} \\ 
\cmidrule(lr){4-6} \cmidrule(lr){7-9}
 & & & A & D & \textbf{Ours} & A & D & \textbf{Ours} \\ 
\midrule
\multirow{4}{*}{SASRec} 
 & \multirow{2}{*}{GPT-5} 
 & NDCG@5 & 0.3528 & 0.3610 & \textbf{0.3865} & 0.2063 & 0.2508 & \textbf{0.2600} \\
 & & HR@5 & 0.4623 & 0.4870 & \textbf{0.5274} & 0.2891 & 0.3427 & \textbf{0.3591} \\
\cmidrule(lr){2-9}
 & \multirow{2}{*}{Claude} 
 & NDCG@5 & 0.3502  & 0.3659& \textbf{0.3913} & 0.2176 & 0.2470 & \textbf{0.2562}  \\
 & & HR@5 & 0.4542 & 0.4901  & \textbf{0.5170} & 0.3056 & 0.3394 & \textbf{0.3578}  \\
\midrule
\multirow{4}{*}{NCF} 
 & \multirow{2}{*}{GPT-5} 
 & NDCG@5 & 0.2556 & 0.2421 & \textbf{0.2723} & 0.1183 & 0.1817 & \textbf{0.1907}  \\
 & & HR@5 & 0.3477 & 0.3497 & \textbf{0.3799} & 0.1733 & 0.2600 & \textbf{0.2714} \\
\cmidrule(lr){2-9}
 & \multirow{2}{*}{Claude} 
 & NDCG@5 & 0.2485 & 0.2478 & \textbf{0.2709} & 0.1218 & 0.1854 & \textbf{0.2018} \\
 & & HR@5 & 0.3311  & 0.3468 & \textbf{0.3724} & 0.1844 & 0.2652 & \textbf{0.2844} \\
\bottomrule[1.5pt]
\end{tabular}}
\vspace{-1ex}
\label{tab: different llm}
\end{wraptable}

To verify the robustness of~\proposed~across different LLMs, we conducted an additional experiment utilizing Claude Opus 4.6 instead of the GPT-5 series used in our main pipeline. From the results in Table~\ref{tab: different llm}, we have the following observations:
\textbf{1)} When utilizing Claude as the backbone LLM, ~\proposed~consistently outperforms the baselines (AlphaEvolve and DeepEvolve). This demonstrates that our core methodology of identifying and resolving structural deficiencies via the User Simulator and Model Diagnosis Tool is fundamentally effective.
\textbf{2)} Furthermore, we observe negligible performance difference between the GPT- and Claude-based variants of ~\proposed. This confirms that our evolutionary paradigm, which is grounded in diagnosing the root causes of model failure, is highly robust and genuinely LLM-agnostic, rather than being an artifact of a specific model's capabilities.


\begin{table}[t]
    \centering
    \caption{Comparison of average execution time per iteration on CDs dataset (Seed Recommender: SASRec). }
    \label{tab:time_cost}
    \resizebox{0.75\linewidth}{!}{
    \begin{tabular}{lcccccr}
        \toprule
        \textbf{Method} & \textbf{RAG} & \textbf{Coding} & \textbf{Co-Evolve} & \textbf{\text{SIM}}  & \textbf{\text{DIAG}} & \textbf{Total Time} \\
        \midrule
        AlphaEvolve & - & 6m 23s & - & - & - & 6m 23s \\
        DeepEvolve & 6m 16s & 7m 43s & - & - & - & 13m 59s \\
        \textbf{Ours} & 6m 24s & 7m 15s & 4m 46s & 44s$^*$ & 12s &19m 21s \\
        \bottomrule
    \end{tabular}
    }
    \begin{flushleft}
        \footnotesize
        $^*$ The User Simulation time is measured with 20 sampled users, processed in parallel. Diagnosis Co-Evolve includes both research (2m 12s) and coding (2m 34s).
    \end{flushleft}
\vspace{-4ex}
\end{table}

\subsection{Efficiency Analysis}\label{app: eff analysis}
\subsubsection{Time Efficiency Analysis}\label{app: time cost}
We compare the execution time per iteration of evolution in Table~\ref{tab:time_cost}. The reported times are averaged over the evolution of SASRec on the CDs dataset. We have the following observations: \textbf{1)} Although~\proposed~requires approximately $19$ minutes per iteration—higher than AlphaEvolve ($6$m) and DeepEvolve ($14$m)—it significantly reduces the \textit{total number of iterations} required to reach peak performance. As illustrated in Table~\ref{tab:initial_model},~\proposed~reaches its peak in just around 8 to 11 iterations, whereas baselines relying on aimless trial-and-error require 13 to 19 iterations or fail to outperform the initial model. {This substantial reduction in required iterations ensures that \proposed~reaches its peak performance highly efficiently. For instance, when evaluated on the random recommender setting for the Electronics dataset, \proposed~achieves peak performance in 3h 27m, significantly outpacing DeepEvolve (4h 19m). Similarly, for SASRec on the Office dataset, \proposed~takes only 1h 54m to reach its peak compared to DeepEvolve's 3h 47m. In both cases, while AlphaEvolve is faster (2h 10m and 1h 32m, respectively), it yields substantially lower performance.} \textbf{2)} While integrating RAG processes in DeepEvolve and~\proposed~increases the runtime per iteration compared to AlphaEvolve, it is critical for ensuring the quality of the evolved logic. As discussed in Sec.~\ref{sec: codebase quality}, the absence of external knowledge limits AlphaEvolve's ability to discover novel mechanisms, resulting in lower \textit{Creativity} and \textit{Insight}. Conversely,~\proposed~leverages this RAG latency to incorporate external algorithmic knowledge and diagnostic signals, enabling the discovery of novel mechanisms that scalar-driven baselines fail to achieve. \textbf{3)} The User Simulator introduces an additional time cost ($44$s), yet it is crucial for sustaining high performance. Furthermore, as detailed in App.~\ref{app: user sample}, the simulator remains stable and effective even with a small number of sampled users, ensuring efficiency without compromising robustness.

\vspace{-1ex}
\subsubsection{Impact of user sampling size.}\label{app: user sample} To investigate the efficiency and robustness of~\proposed~on $\mathcal{U}_{\text{sample}}$, we conducted experiments by varying the user sample size of the User Simulator. As shown in Figure~\ref{fig:sample_size}, we observe that~\proposed~achieves robust performance even with a small number of sampled users. Notably, performance improves significantly as the sample size increases from 1 to 3, eventually stabilizing around a sample size of 5. This robustness stems from the \textbf{Diagnosis Tool - Model Co-Evolution} mechanism (Sec.~\ref{Sec Co-evolution}). Although the \text{SIM} operates on a small subset of users to generate qualitative feedback (e.g., "lack of diversity"), the \text{DIAG} translates these critiques into numerically-grounded metrics (e.g., measuring category entropy). Consequently, even with limited user samples, the evolved \text{DIAG} effectively verifies and quantifies structural deficiencies across the global data distribution, ensuring reliable evolutionary guidance without the need for extensive user sampling.

\vspace{-.5ex}
\subsubsection{Impact of the size of literature.}\label{app: literature sample}
To investigate the effect of literature size (i.e., $\mathcal{K}$) on~\proposed, we vary the number of literature utilized in the Planning (Section~\ref{Sec Evolution Pipeline}). As shown in Figure~\ref{fig:literature_sample}, we observe that~\proposed~maintains robust performance across different size of $\mathcal{K}$. This robustness stems from the directional feedback provided by~\proposed, which effectively pinpoints issues within the recommender pipeline. Consequently, even without searching a massive amount of external knowledge,~\proposed~can accurately diagnose and correct the problems in the current recommender.

\vspace{1ex}
\subsubsection{Efficiency Analysis under Fixed Token Budget}
\begin{wraptable}{r}{0.5\textwidth}
\vspace{-2.5ex}
\caption{Performance comparison under a fixed token budget (Seed Recommender: SASRec).}
\label{tab:budget}
\resizebox{0.95\linewidth}{!}{%
\begin{tabular}{ccccc}
\toprule[1.5pt]
\multirow{2}{*}{\textbf{Method}} & \multicolumn{2}{c}{\textbf{CDs}} & \multicolumn{2}{c}{\textbf{Electronics}} \\ 
\cmidrule(lr){2-3} \cmidrule(lr){4-5}
 & NDCG@5 & HR@5 & NDCG@5 & HR@5 \\ 
\midrule
AlphaEvolve   & 0.3528 & 0.4623 & 0.1995 & 0.2785  \\
DeepEvolve   & 0.3539 & 0.4809 & 0.2415 & 0.3352 \\
\proposed   & \textbf{0.3843} & \textbf{0.5209} & \textbf{0.2472} & \textbf{0.3580} \\
\bottomrule[1.5pt]
\vspace{-2ex}
\end{tabular}%
}
\end{wraptable}
To ensure a fair comparison regarding computational costs, we evaluate all evolutionary frameworks (i.e., AlphaEvolve, DeepEvolve, and~\proposed) under a fixed token budget. Specifically, we constrain the total number of tokens (including both input prompts and completions) utilized throughout the entire evolutionary process to 5M tokens.
As shown in Table~\ref{tab:budget}, we observe that~\proposed~consistently outperforms the baselines even under a fixed token budget. This superiority stems from the ability of~\proposed~to leverage directional feedback to pinpoint specific failures within the recommendation pipeline and effectively resolve them. Consequently,~\proposed~achieves more effective evolution within the same resource constraints by avoiding the aimless trial-and-error search characteristic of scalar-driven methods.

\vspace{-.5ex}
\subsubsection{Efficiency via Zero-Cost Proxies}\label{app: zero cost eff}
\begin{wraptable}{r}{0.475\textwidth}
\vspace{-6.5ex}
\centering
\captionsetup{font=small}
\caption{Performance and efficiency of zero-cost proxy variant. Time denotes average duration per iteration for training and directional feedback generation.}
\vspace{-1ex}
\setlength{\tabcolsep}{4pt}
\resizebox{0.925\linewidth}{!}{%
\begin{tabular}{cccccc}
\toprule[1.5pt]
Seed & Dataset & Type & NDCG@5 & HR@5 & Time \\
\midrule
\multirow{4}{*}{SASRec} 
 & \multirow{2}{*}{CDs}  & Original  & \textbf{0.3865} & \textbf{0.5274} & 2m 58s \\
 &                        & Zero-Cost & 0.3798 & 0.5054 & \textbf{36s} \\
\cmidrule(lr){2-6}
 & \multirow{2}{*}{Elec.} & Original  & \textbf{0.2600} & \textbf{0.3591} & 3m 54s \\
 &                        & Zero-Cost & 0.2531 & 0.3485 & \textbf{39s} \\
\midrule
\multirow{4}{*}{MoRec} 
 & \multirow{2}{*}{CDs}  & Original  & \textbf{0.3977} & \textbf{0.5340} & 12m 08s \\
 &                        & Zero-Cost & 0.3827 & 0.4991 & \textbf{1m 02s} \\
\cmidrule(lr){2-6}
 & \multirow{2}{*}{Elec.} & Original  & \textbf{0.2056} & \textbf{0.2921} & 7m 10s \\
 &                        & Zero-Cost & 0.1964 & 0.2869 & \textbf{1m 13s} \\
\bottomrule[1.5pt]
\end{tabular}}
\label{tab: zero cost proxy}
\end{wraptable}
To validate the efficiency of the variant in Sec.~\ref{Sec efficient variant}, we compare the zero-cost proxy variant of~\proposed~against its fully-trained pipeline in Table~\ref{tab: zero cost proxy}. We have the following observations: \textbf{1)} The zero-cost proxy variant achieves comparable recommendation performance to the fully-trained pipeline, with only marginal degradation. \textbf{2)} The variant substantially reduces the per-iteration computational overhead, achieving up to 12$\times$ speedup (e.g., 12m 08s $\to$ 1m 02s on MoRec/CDs). These results confirm that the efficient variant offers a practical trade-off, retaining most of the recommendation quality while drastically reducing the computational cost of evolution.

\begin{table}[h]
\centering
\caption{Recommendation performance under different persona schemes for the User Simulator (Seed Recommender: SASRec).}
\label{tab:big 5}
\resizebox{0.8\linewidth}{!}{%
\begin{tabular}{ccccc}
\toprule[1.5pt]
\multirow{2}{*}{\textbf{Persona Scheme}} & \multicolumn{2}{c}{\textbf{CDs}} & \multicolumn{2}{c}{\textbf{Electronics}} \\ 
\cmidrule(lr){2-3} \cmidrule(lr){4-5}
 & NDCG@5 & HR@5 & NDCG@5 & HR@5 \\ 
\midrule
Activity / Conformity / Diversity (Ours) & 0.3865 & \textbf{0.5274} & \textbf{0.2600} & \textbf{0.3591} \\ 
Big Five  & \textbf{0.3915} & 0.5244 & 0.2551 & 0.3561 \\
\bottomrule[1.5pt]
\end{tabular}%
}
\vspace{-2ex}
\end{table}

\subsection{Robustness to Persona Scheme}\label{exp big five}
To verify that \proposed~is not tied to a specific user characterization, we replace the default traits (Activity, Conformity, Diversity) with the Big Five personality traits (Openness, Conscientiousness, Extraversion, Agreeableness, and Neuroticism), following prior work~\cite{ma2025pub}. As shown in Table~\ref{tab:big 5}, \proposed~achieves comparable recommendation performance under both persona schemes. This is consistent with our observation in Sec.~\ref{exp sim evolve} that SIM already captures complex user preferences reliably, and indicates that the driving factor of evolution is the precise, qualitative nature of $\mathcal{R}_{\text{SIM}}$ rather than the specific trait taxonomy used to construct it. These results confirm that \proposed~is robust to variations in user characterization.



\subsection{{Hallucinations and Groundedness of the User Simulator}}
\label{app: hallucination groundedness}

To examine whether the User Simulator produces factual hallucinations, such as fabricated items, interactions, or user preferences, we analyzed all simulation feedback generated along the evolutionary trajectories in Table~\ref{tab:overall_performance}. However, we found no cases in which the simulator made claims contradicted by the actual user histories or recommendation lists.

To further quantify this groundedness, we adopt the Interest Groundedness (IG) metric from GISTBench~\citep{fostiropoulos2026gistbench}, which evaluates whether an LLM-inferred user interest is supported by evidence from the user's interaction history, rather than merely sounds plausible. Following GISTBench, we define $\mathcal{I}_{pred}$ as the set of interest categories inferred from the simulator's feedback (e.g., item title or category preference), and $\mathcal{I}_{true}$ as the set verified from the user's historical interactions. Since our recommendation datasets contain relatively sparse user histories, we treat an interest category as verified when it is supported by at least one explicit positive interaction with a rating of 4 or higher.

We compute IG precision and recall as:
\begin{equation}
\footnotesize
IG_{Precision} = \frac{|\mathcal{I}_{pred} \cap \mathcal{I}_{true}|}{|\mathcal{I}_{pred}|}, \quad
IG_{Recall} = \frac{|\mathcal{I}_{pred} \cap \mathcal{I}_{true}|}{|\mathcal{I}_{true}|}
\end{equation}
{Following GISTBench, all metrics are computed per user and aggregated as the median across users for each evolutionary trajectory, and then averaged across trajectories.}
Here, $IG_{Precision}$ directly captures the absence of hallucinations by measuring whether the simulator's inferred interests are grounded in evidence from the user's history, while $IG_{Recall}$ measures the coverage of interests verifiable from the user's history.


\begin{wraptable}{r}{0.25\textwidth}
\vspace{-2ex}
\centering
\caption{Performance of Interest Groundedness.}
\label{tab:reliability}
\scriptsize
\setlength{\tabcolsep}{1pt}
\resizebox{1.0\linewidth}{!}{
\begin{tabular}{lcc}
\toprule
Dataset & $IG_{Precision}$ & $IG_{Recall}$ \\
\midrule
CDs & 0.891 & 0.749\\
Electronics & 0.843 & 0.578 \\
Office & 0.851 & 0.570 \\
\bottomrule
\end{tabular}}
\vspace{-2ex}
\label{tab:reliability}
\end{wraptable}
As shown in Table~\ref{tab:reliability}, the simulator achieves high $IG_{Precision}$ on both CDs and Electronics, confirming that nearly all inferred interests are grounded in explicit positive evidence and the simulator does not hallucinate user preferences. The $IG_{Recall}$ scores further show that it covers a substantial portion of verifiable interests, suggesting that simulator feedback is largely grounded in observed user behavior.

\smallskip
\textbf{Complementary Interaction of SIM and DIAG.}

While high IG scores confirm that the simulator is free from factual hallucinations, they do not guarantee that its critiques are always technically precise or aligned with the model's internal state, necessitating calibration via DIAG to distinguish between surface-level complaints and actual structural failures. Specifically, \textbf{1)} the simulator occasionally over-generalizes semantic overlaps. For instance, seeing multiple albums by the same artist (e.g., Adele) might lead the simulator to harshly criticize the model for "severe redundancy," even though each item is a valid historical interest. DIAG recalibrates this by verifying that the duplicate exposure rate is near-zero and items are geometrically distinct, proving the critique is a semantic over-generalization rather than a structural representation collapse. \textbf{2)} The simulator lacks structural awareness. In our trajectory investigation, it often repeated the same semantic complaints even as the model's geometric health (e.g., reduced embedding collapse) was objectively resolving. This confirms that DIAG is essential not just for verifying the absence of hallucinations, but for bridging the gap between the simulator's user-centric perception and the model's actual architectural progress.

\subsection{Additional Case Studies}
\subsubsection{Addtional Case Study: Reliability of Co-evolved Diagnostic Tool via Deficiencies Injection.}\label{App: Case Study Injection}
\begin{wrapfigure}{r}{0.5\textwidth}
    \centering    
    \includegraphics[width=1.0\linewidth]{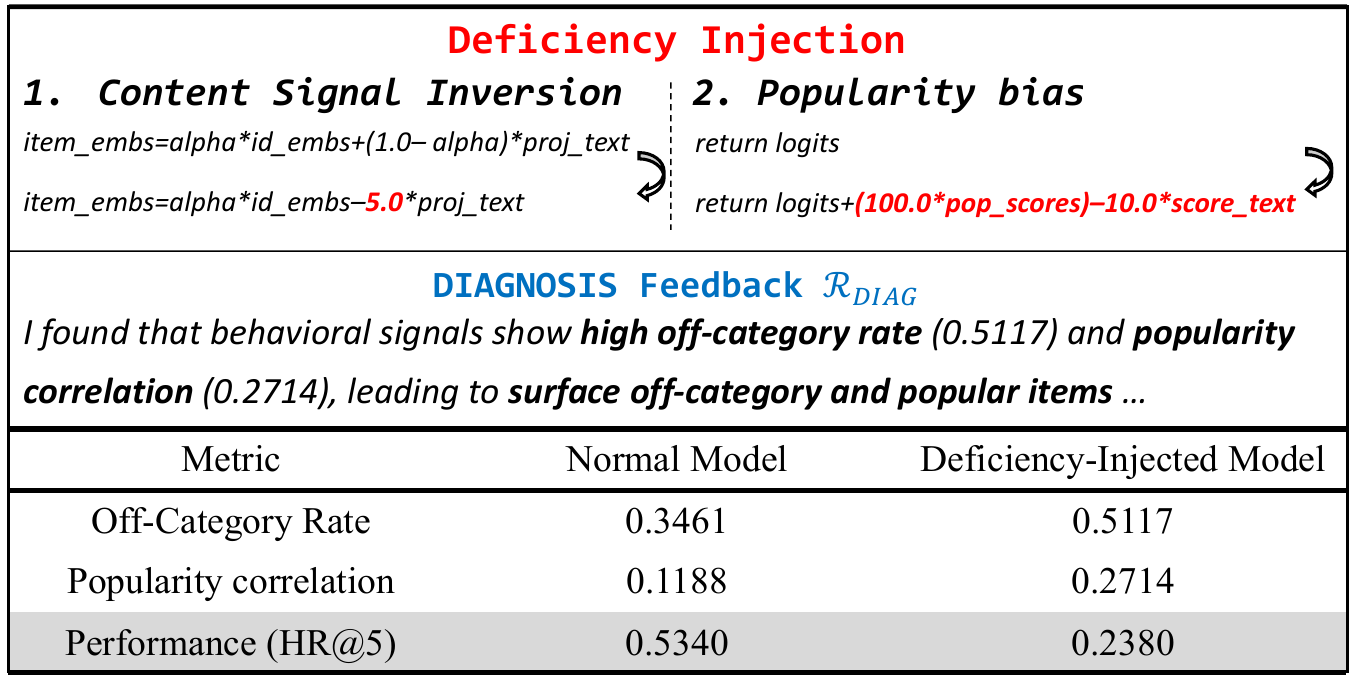}
    \vspace{-1.5ex}
    \captionsetup{font=small}
    \caption
    {Case study on Diagnosis Tool - Model Co-Evolution on CDs dataset (Seed Recommender: MoRec).}
    \vspace{-2ex}
    \label{fig: diagnosis_casestudy_morec}
\end{wrapfigure}
In case Figure~\ref{fig: diagnosis_casestudy_morec}, we injected deficiencies into an evolved MoRec pipeline by: (i) inverting content signals, and (ii) inflating popularity scores. The co-evolved DIAG effectively detects these deficiencies through newly formulated metrics such as the \textbf{Off-Category Rate} (Mismatch rate between the categories of the Top-K recommended items and the categories present in user's recent interaction history) and \textbf{Popularity correlation} (Correlation between the model's predicted recommendation logits and the item popularity distribution). The $\mathcal{R}_{DIAG}$ accurately reports that the system has begun to "surface off-category or popular items rather than reliably relevant ones" providing clear directional feedback for subsequent correction. Notably, the strong association between these detected deficiencies and the performance drop confirms that the co-evolved metrics can effectively bridge the gap between internal behavioral shifts and external recommendation accuracy.

\subsubsection{Additional Case Study: Evolutionary Trajectory.}\label{app: case study trajectory}
\textbf{Evolutionary Trajectory on SASRec: CDs dataset.}

\begin{wrapfigure}{r}{0.425\textwidth}
\centering    \includegraphics[width=0.99\linewidth]{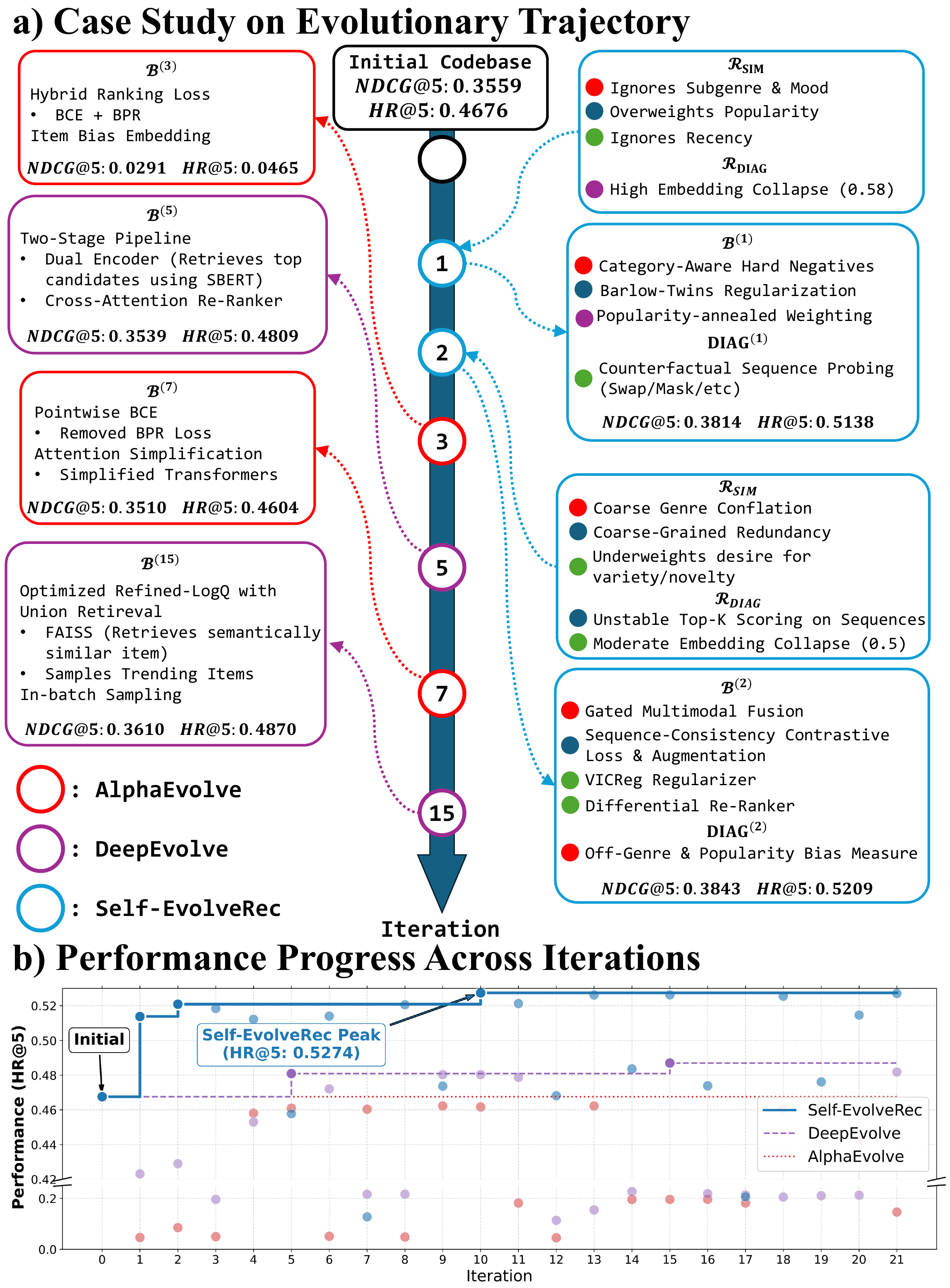}
    \vspace{-4ex}
    \captionsetup{font=small}
    \caption
    {Case study on evolutionary trajectory on CDs dataset (Seed Recommender: SASRec). (a) is comparison of evolutionary paths. Color-coded markers (e.g., Red) illustrate causal alignment between directional feedback and evolved codebase. (b) is performance comparison across iterations.}
    \label{fig: evolve_casestudy}
    \vspace{-1ex}
\end{wrapfigure}

To validate the effectiveness of the directional feedback loop, we conducted a case study comparing the evolutionary trajectories of \proposed~against baselines. Figure~\ref{fig: evolve_casestudy} (a) illustrates the step-by-step evolution of codebases, while Figure~\ref{fig: evolve_casestudy} (b) tracks the performance progress over all iterations.
We have the following observations: \textbf{1)}~\proposed~shows a structured evolutionary path, where algorithmic improvements are causally linked to identified failures. For instance, in Iteration ($0\rightarrow 1$), $\mathcal{R}_{\text{SIM}}$ explicitly flagged "Ignores Subgenre", while $\mathcal{R}_{\text{DIAG}}$ detected "High Embedding Collapse." Guided by this directional feedback, the agent introduced "Category-Aware Hard Negatives" and "Popularity-annealed Weighting," resulting in an immediate performance increment (HR: $0.4676 \rightarrow 0.5138$).
\textbf{2)} In contrast, baselines exhibit unstable or delayed progress due to their reliance on scalar metrics. AlphaEvolve attempts an erroneous combination of loss functions (BCE + BPR) at Iteration 3, causing a significant performance drop, which is only rectified by removing the module at Iteration 7. Consequently, it does not exceed its initial performance state throughout the evolution.
DeepEvolve suffers from prolonged stagnation in low-performance regions, as evident in Figure~\ref{fig: evolve_casestudy} (b), and only manages a gain at Iteration 15 by retrieving the `Refined-LogQ' module. Due to such inefficient exploration, as observed in Figure~\ref{fig: evolve_casestudy} (b), both baselines remain trapped in low-performance regions ($HR@5: 0.05-0.25$), failing to escape suboptimal states throughout the evolution process. Conversely,~\proposed~leverages directional feedback to maintain a robust evolutionary trajectory.

\clearpage
\noindent\textbf{Evolutionary Trajectory on Random Recommender: CDs dataset.}

\begin{wrapfigure}{r}{0.4\textwidth}
\vspace{-2ex}
    \centering    \includegraphics[width=1.0\linewidth]{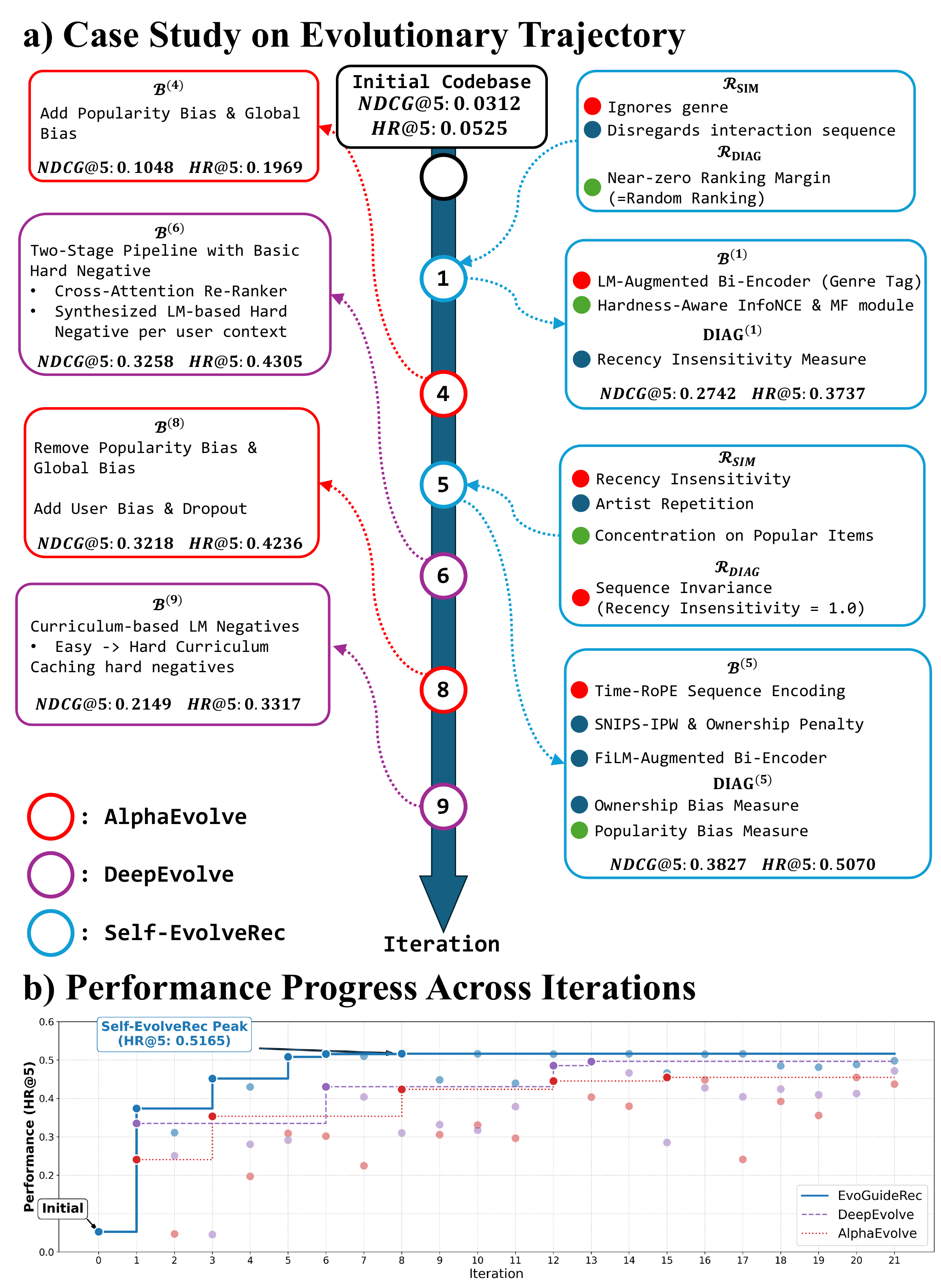}
    \captionsetup{font=small}
    \caption
    {Case study on evolutionary trajectory on CDs dataset (Seed Recommender: Random). (a) is comparison of evolutionary paths. Color-coded markers (e.g., Red) illustrate causal alignment between directional feedback and evolved codebase. (b) is performance comparison across iterations.}
    \label{fig: evolve_casestudy_appendix}
\end{wrapfigure}
To investigate the~\proposed's behavior on extreme initial recommender setting, we examined \proposed's behavior starting from a Random Recommender (Sec.~\ref{exp: extreme initial}). Figure~\ref{fig: evolve_casestudy_appendix}(a) illustrates the step-by-step code evolution, while Figure~\ref{fig: evolve_casestudy_appendix}(b) tracks the performance progress. We have the following observations:
\textbf{1)}~\proposed~demonstrates a structured evolutionary path where algorithmic improvements are causally linked to identified failures. For instance, in the transition from Iteration $0 \rightarrow 1$, $\mathcal{R}_{\text{DIAG}}$ explicitly flagged the "Random Ranking" behavior, while $\mathcal{R}_{\text{SIM}}$ highlighted the neglect of item categories ("Ignores genre"). Guided by this directional feedback,~\proposed~introduced an "LM-Augmented Bi-Encoder" to embed "Genre Tags" and utilized the InfoNCE loss with an MF module. This effectively resolved the random recommendation issue, yielding a significant performance leap (HR: $0.0525 \rightarrow 0.3737$). Similarly, at Iteration 5,~\proposed~detected "Recency Insensitivity" and addressed it by integrating "Time-RoPE Sequence Encoding," further boosting performance to HR: 0.5070. \textbf{2)} Consistent with the findings in evolutionary trajectory on SASRec, baselines exhibit unstable or delayed progress due to their reliance on scalar metrics without diagnostic guidance. At Iteration 4, AlphaEvolve attempted to add "Popularity Bias and Global Bias" to the model, but this update degraded performance (HR:2405 $\rightarrow$ 0.1969), as depicted in Figure~\ref{fig: evolve_casestudy_appendix} (b). Consequently, AlphaEvolve removes these changes and add "User Bias" at iteration 8, illustraing the inefficient tiral-and-error process.
DeepEvolve shows a successful evolution with a "Two-Stage Pipeline" at iteration 6, but failed to improve at iteration 9 due to an incompatible curriculum learning strategy for LM negatives. Also in Figure~\ref{fig: evolve_casestudy_appendix} (b) confirms that while baselines suffer from performance fluctuations,~\proposed~maintains a robust evolutionary trajectory enabled by directional feedback.

\section{Prompts}
\label{app prompts}
\subsection{Task-specific Instruction Prompts}
\label{app prompts task}
The instruction prompts for the code evolution process were formulated by drawing upon existing methodologies~\cite{novikov2025alphaevolve, liu2025scientific}, ensuring consistency with established benchmarks.

\clearpage
\begin{figure*}[h]
  \centering
  \includegraphics[width=0.8\linewidth]{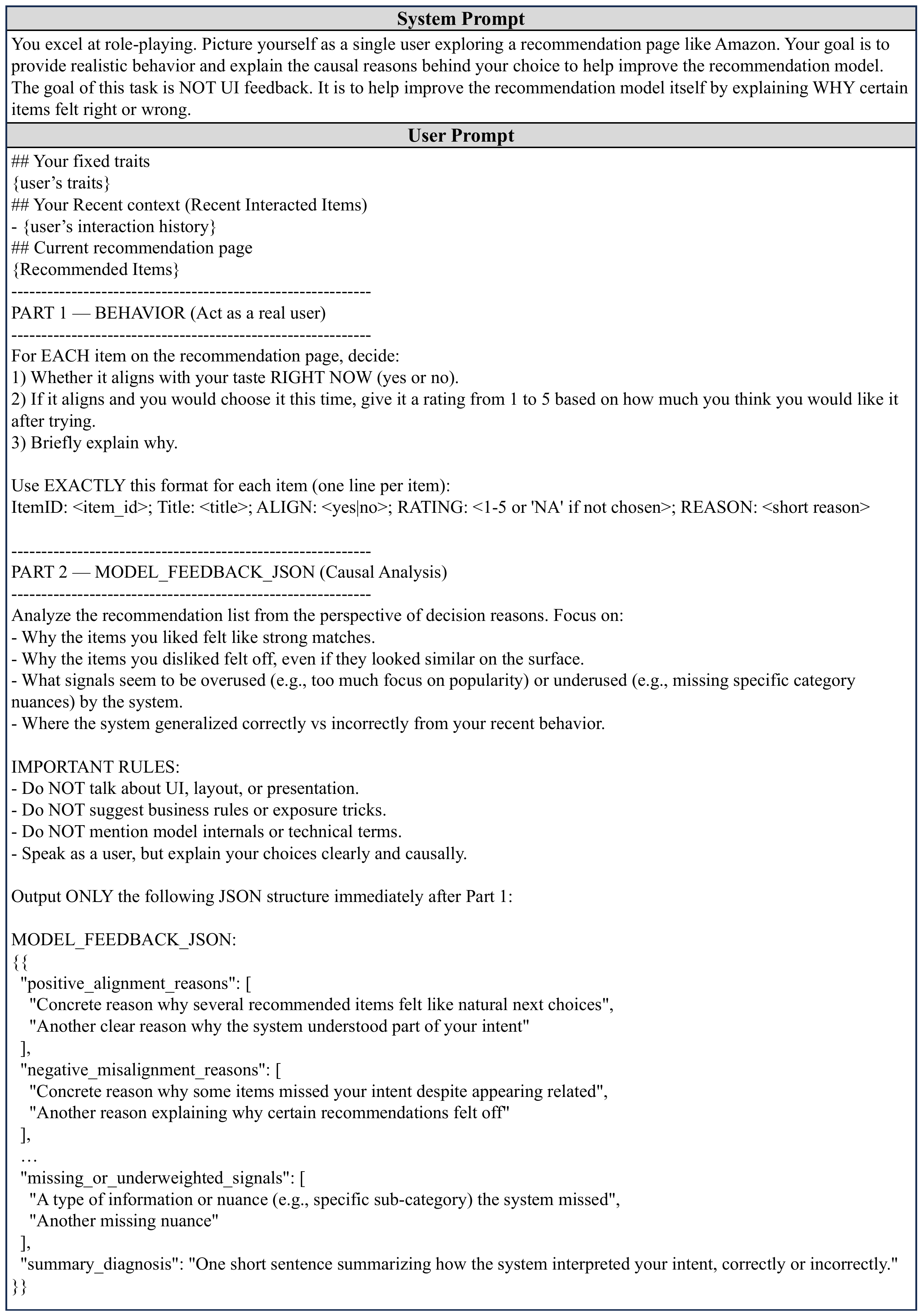}
  \caption
    {Example prompt of $\mathcal{I}_{\text{SIM}}$.}
    \label{fig: I_sim}
\end{figure*}

\begin{figure*}[h]
  \centering
  \includegraphics[width=0.8\linewidth]{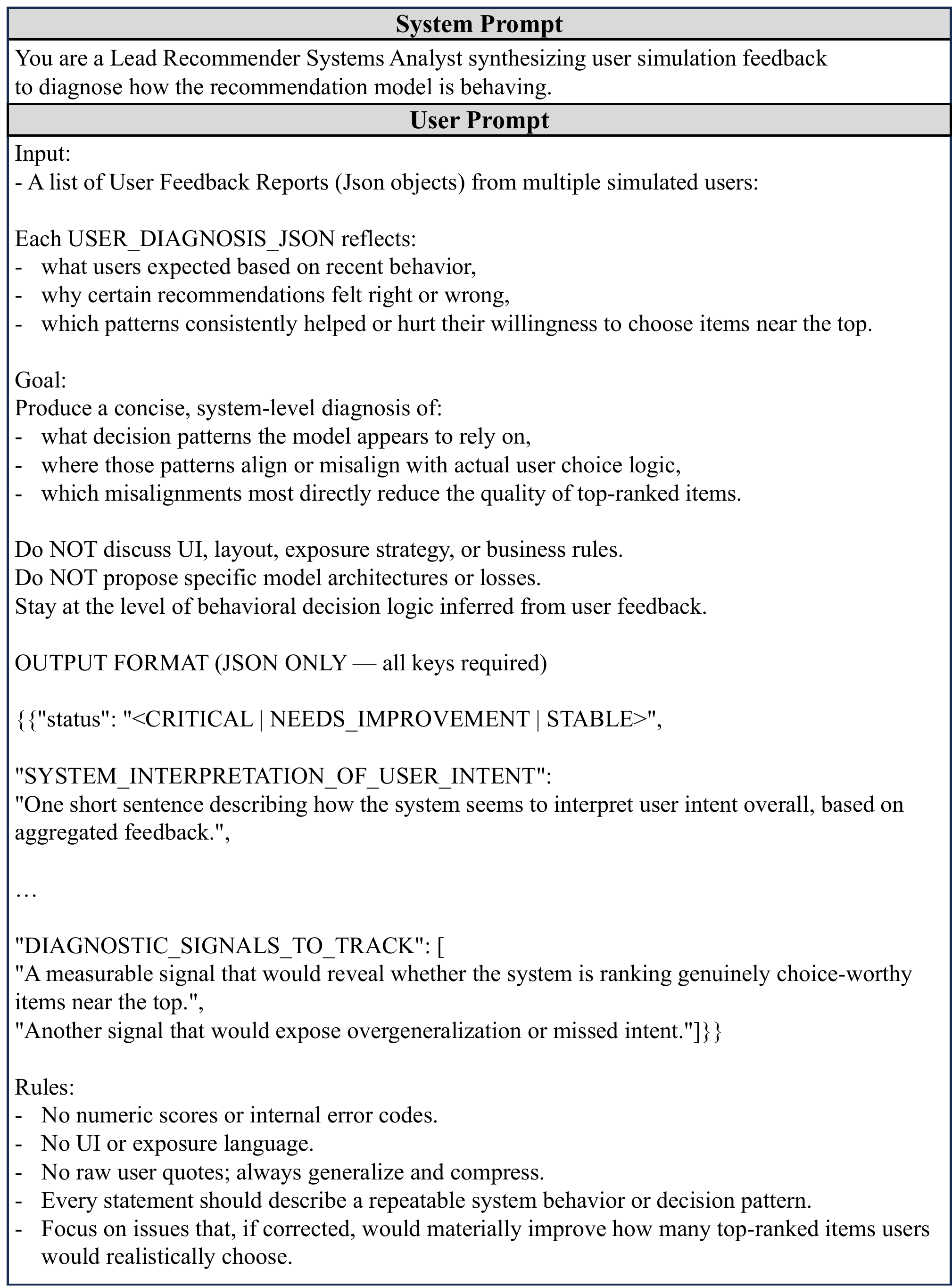}
  \caption
    {Example prompt of $\mathcal{I}_{\text{SUMMARIZE}}$.}
    \label{fig: I_summarize}
\end{figure*}

\begin{figure*}[h]
  \centering
  \includegraphics[width=0.8\linewidth]{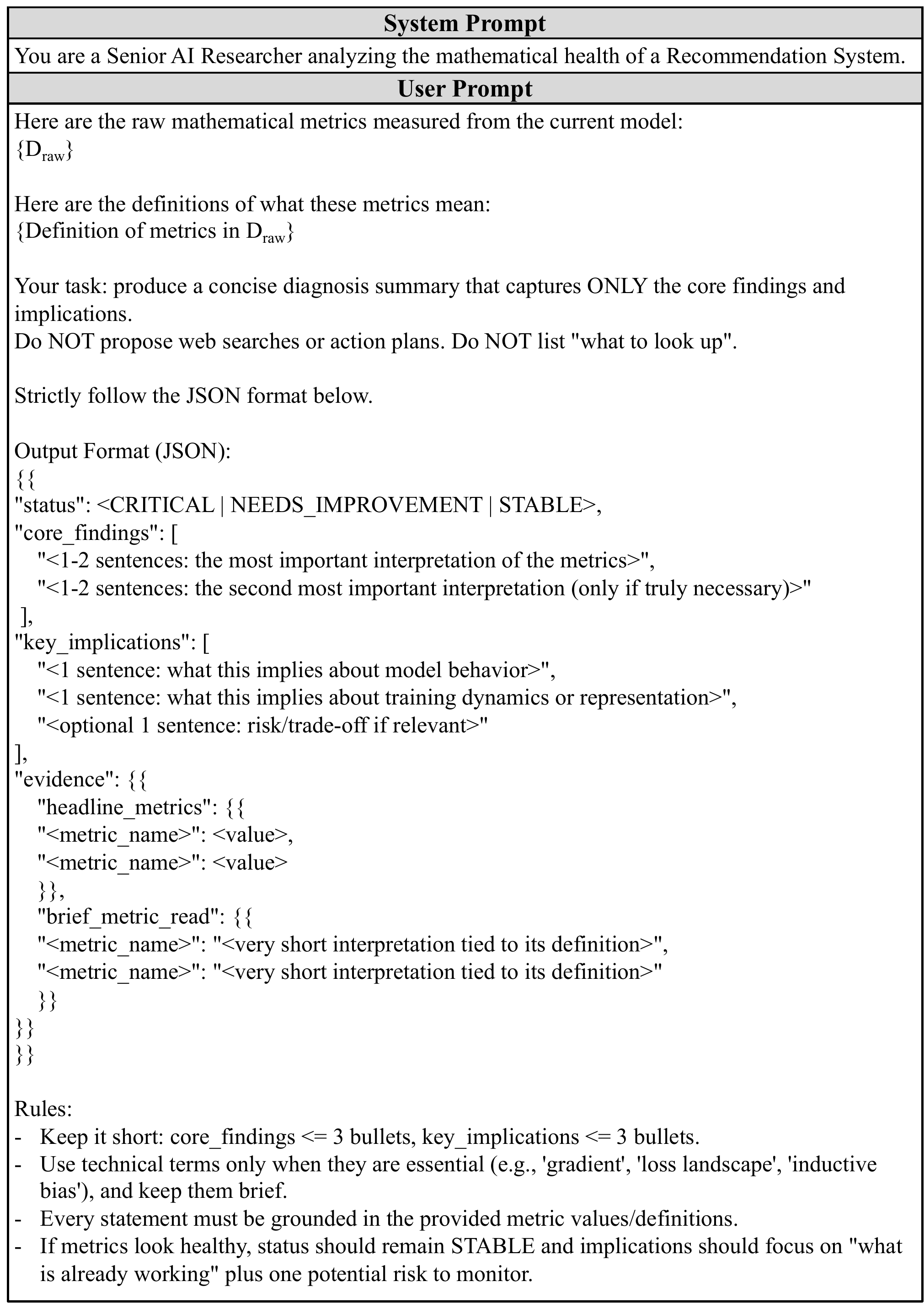}
  \caption
    {Example prompt of $\mathcal{I}_{\text{DIAG}}$.}
    \label{fig: I_diag}
\end{figure*}

\begin{figure*}[h]
  \centering
  \includegraphics[width=0.8\linewidth]{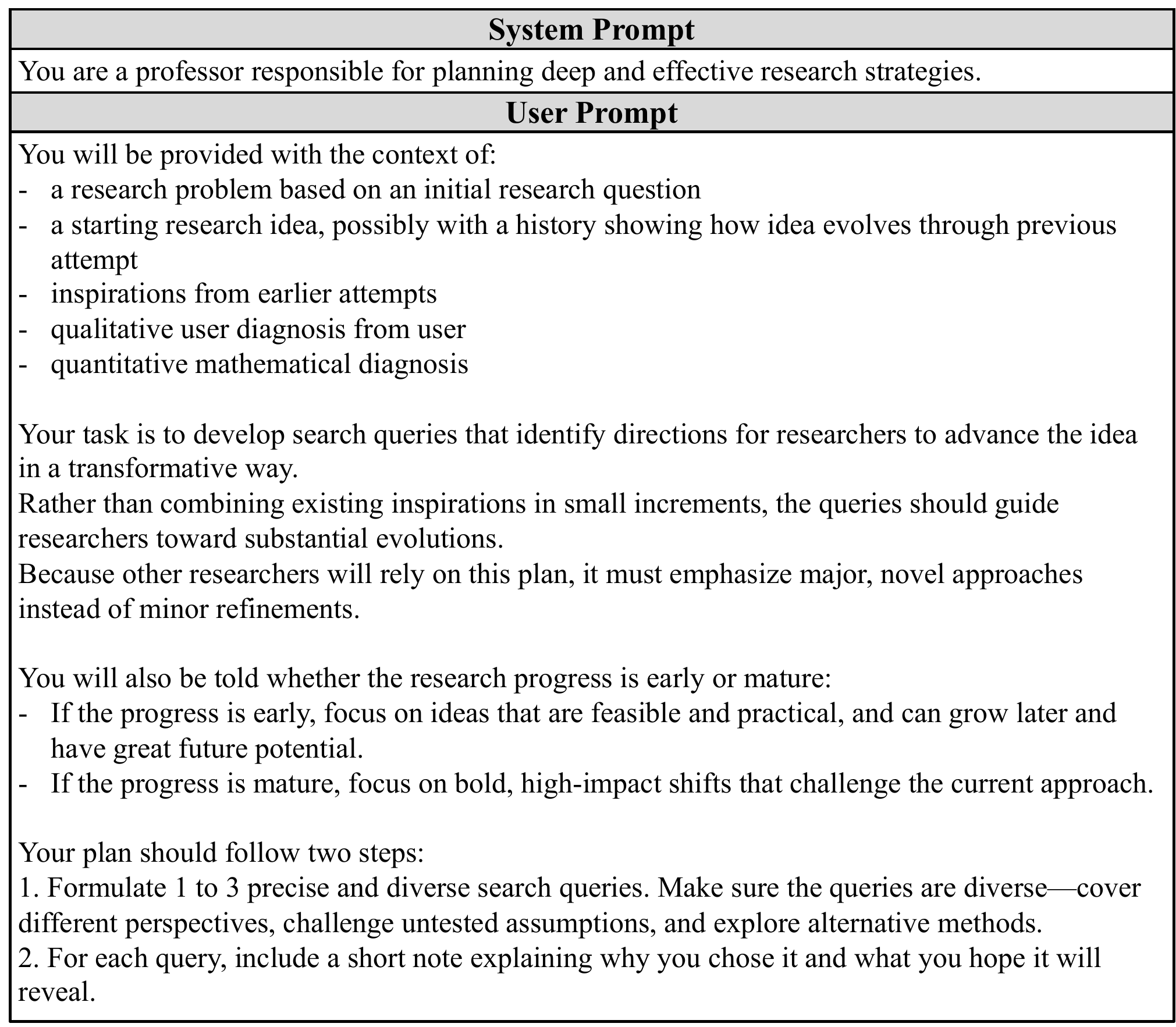}
  \caption
    {Example prompt of $\mathcal{I}_{\text{PLAN}}$.}
    \label{fig: I_plan}
\end{figure*}

\begin{figure*}[h]
  \centering
  \includegraphics[width=0.8\linewidth]{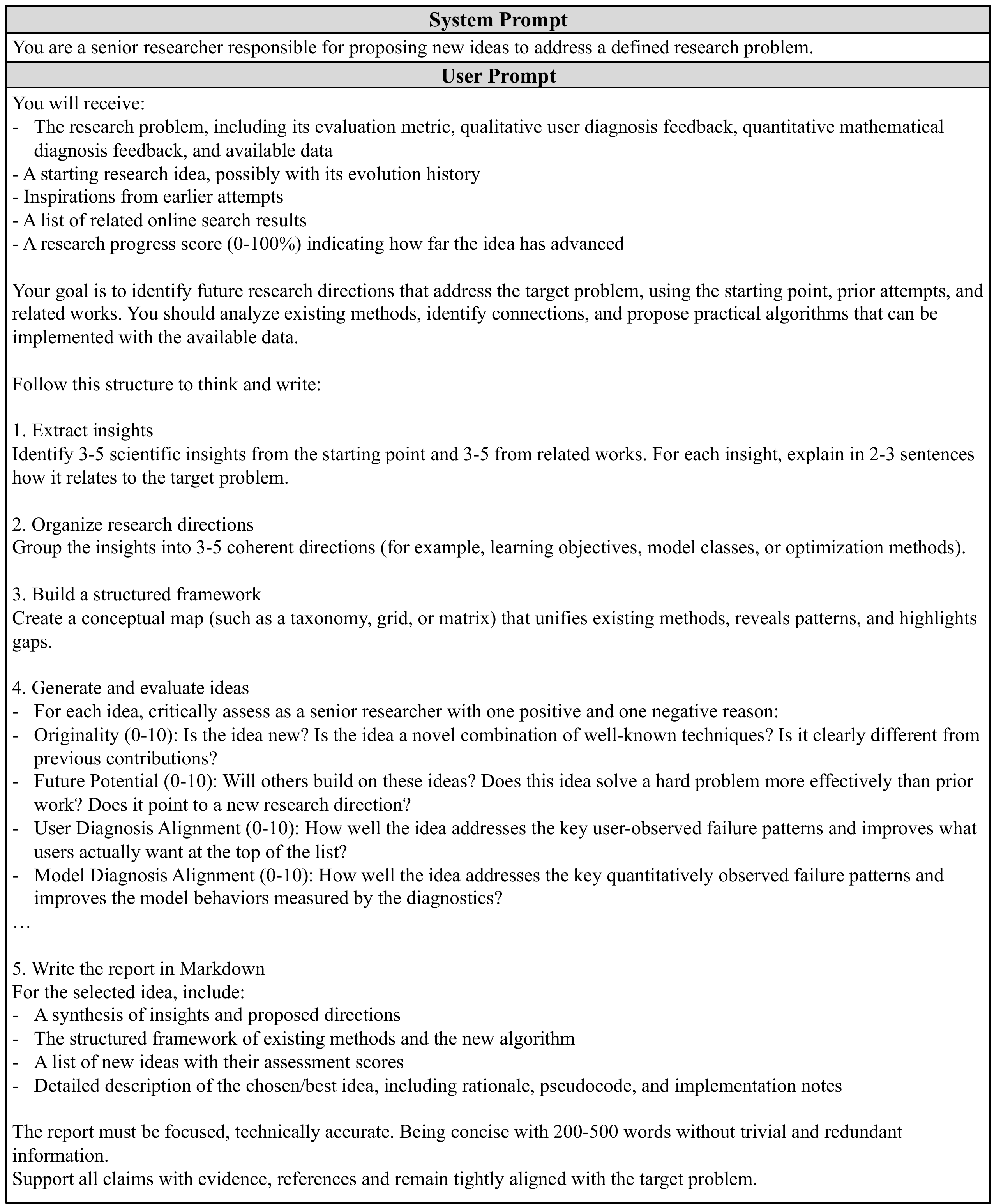}
  \caption
    {Example prompt of $\mathcal{I}_{\text{REPORT}}$.}
    \label{fig: I_report}
\end{figure*}

\begin{figure*}[t]
  \centering
  \includegraphics[width=0.8\linewidth]{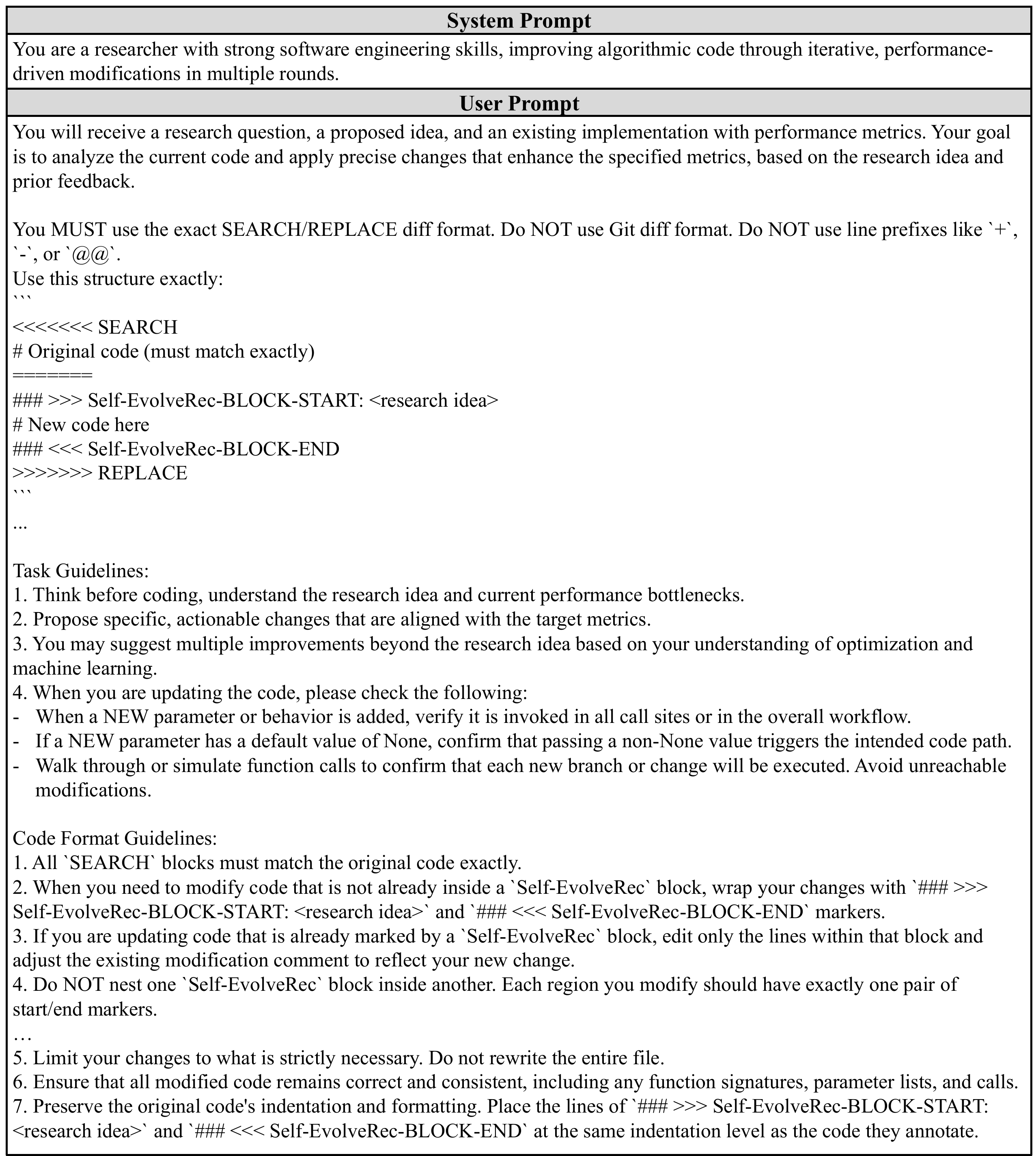}
  \caption
    {Example prompt of $\mathcal{I}_{\text{CODE}}$.}
    \label{fig: I_code}
\end{figure*}

\begin{figure*}[h]
  \centering
  \includegraphics[width=0.8\linewidth]{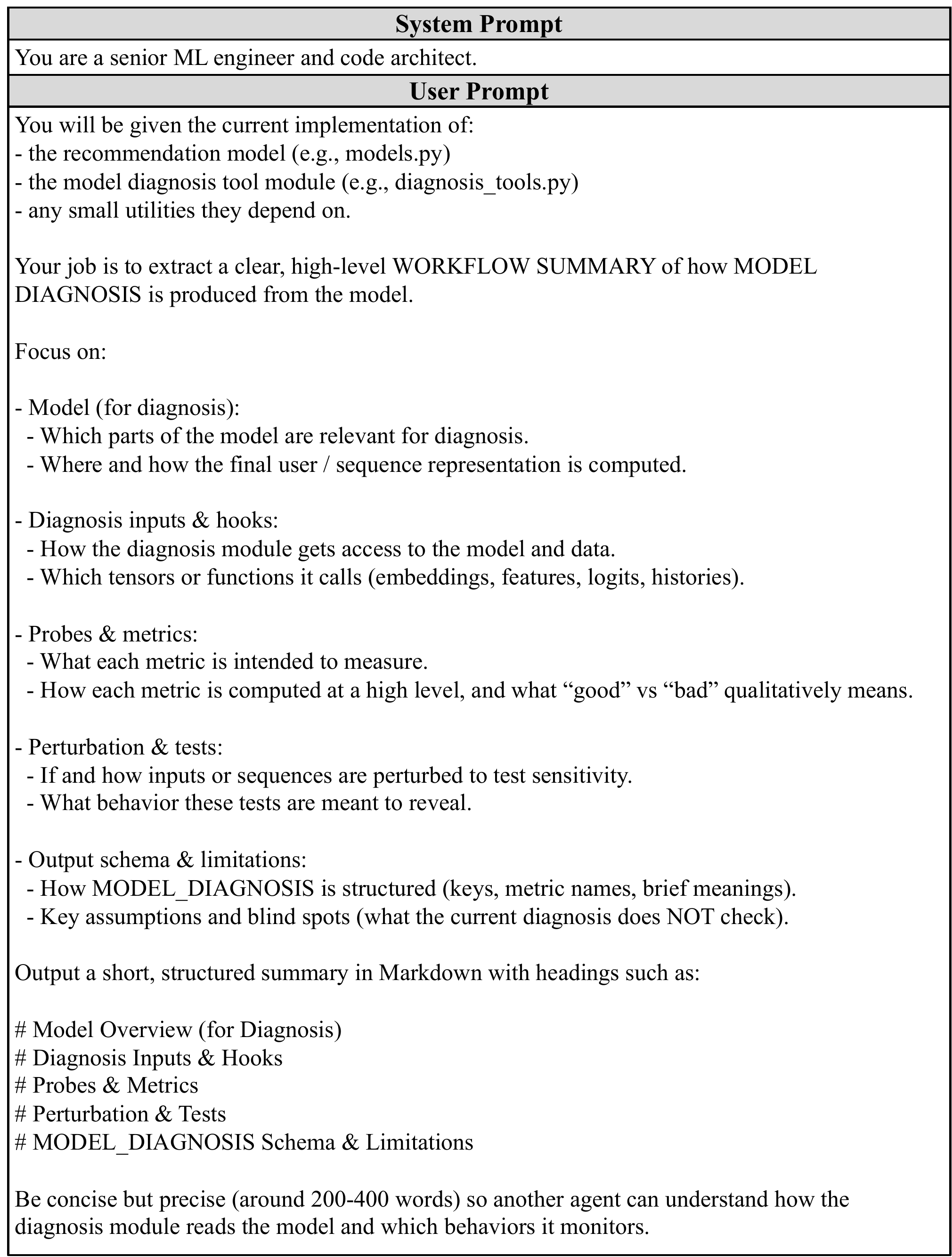}
  \caption
    {Example prompt of $\mathcal{I}_{\text{Analyze}}$.}
    \label{fig: I_analyze}
\end{figure*}

\begin{figure*}[h]
  \centering
  \includegraphics[width=0.8\linewidth]{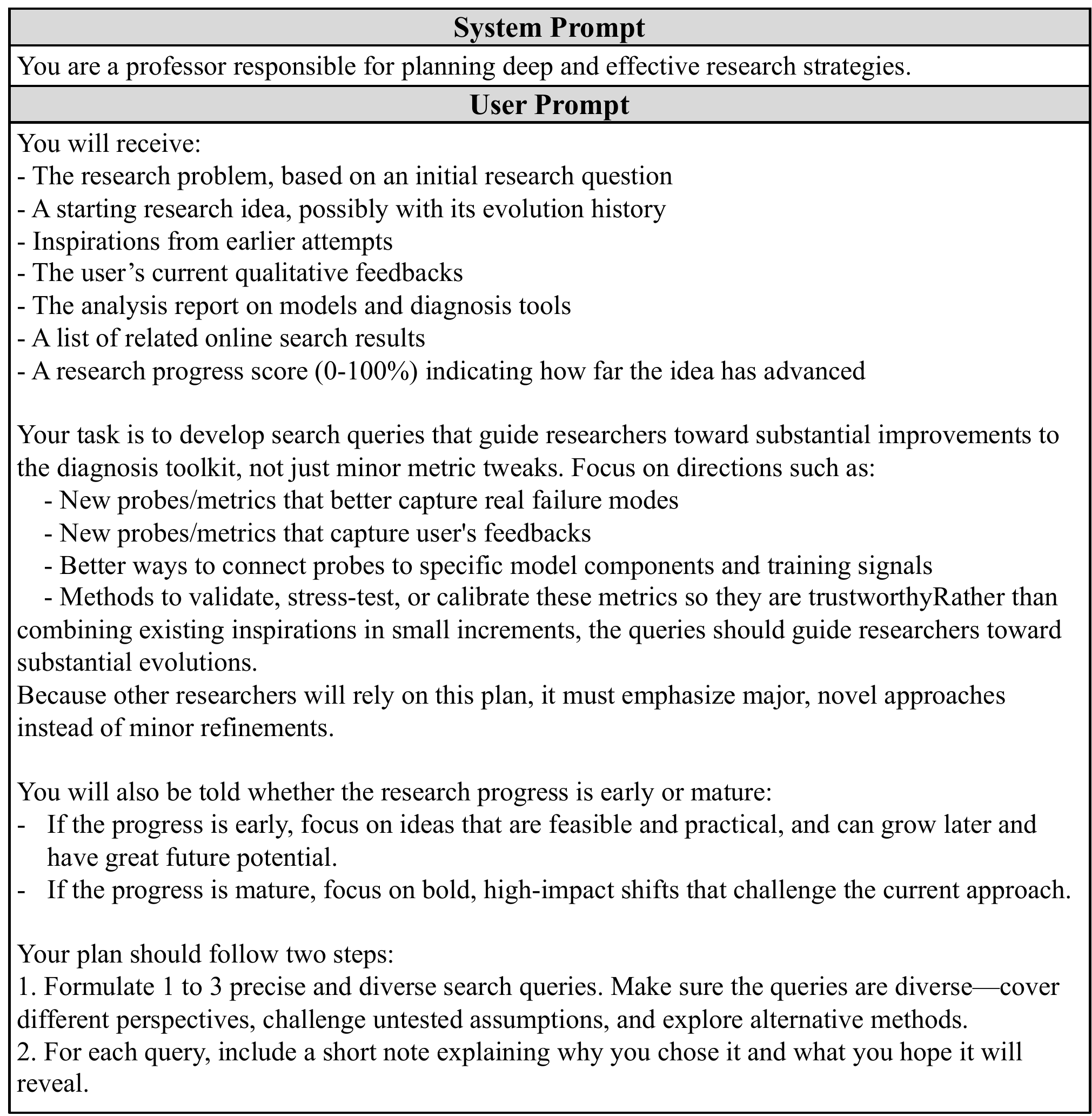}
  \caption
    {Example prompt of $\mathcal{I}_{\text{PLAN-DIAG}}$.}
    \label{fig: I_plan_diag}
\end{figure*}

\begin{figure*}[h]
  \centering
  \includegraphics[width=0.8\linewidth]{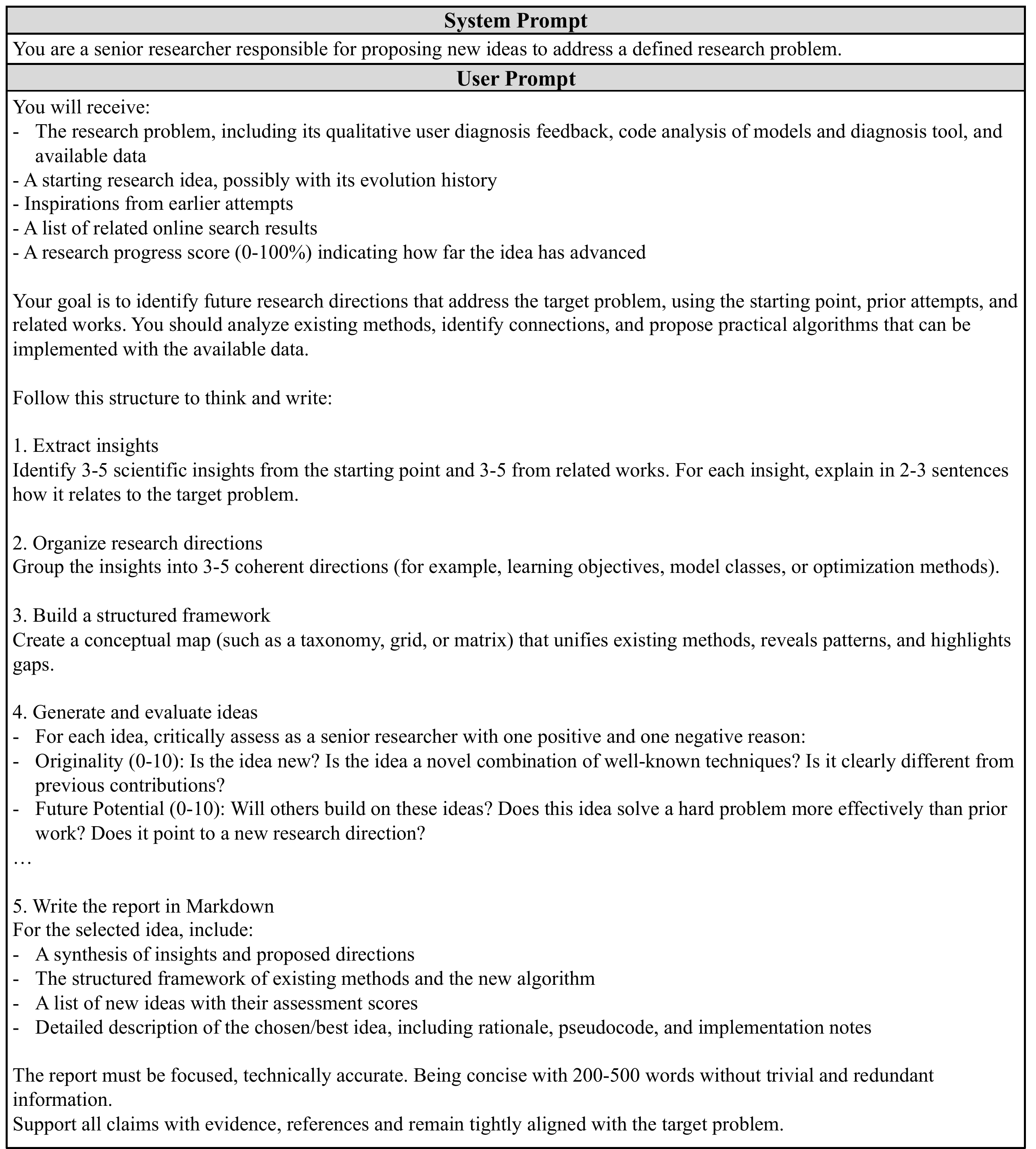}
  \caption
    {Example prompt of $\mathcal{I}_{\text{REPORT-DIAG}}$.}
    \label{fig: I_report-diag}
\end{figure*}

\begin{figure*}[h]
  \centering
  \includegraphics[width=0.8\linewidth]{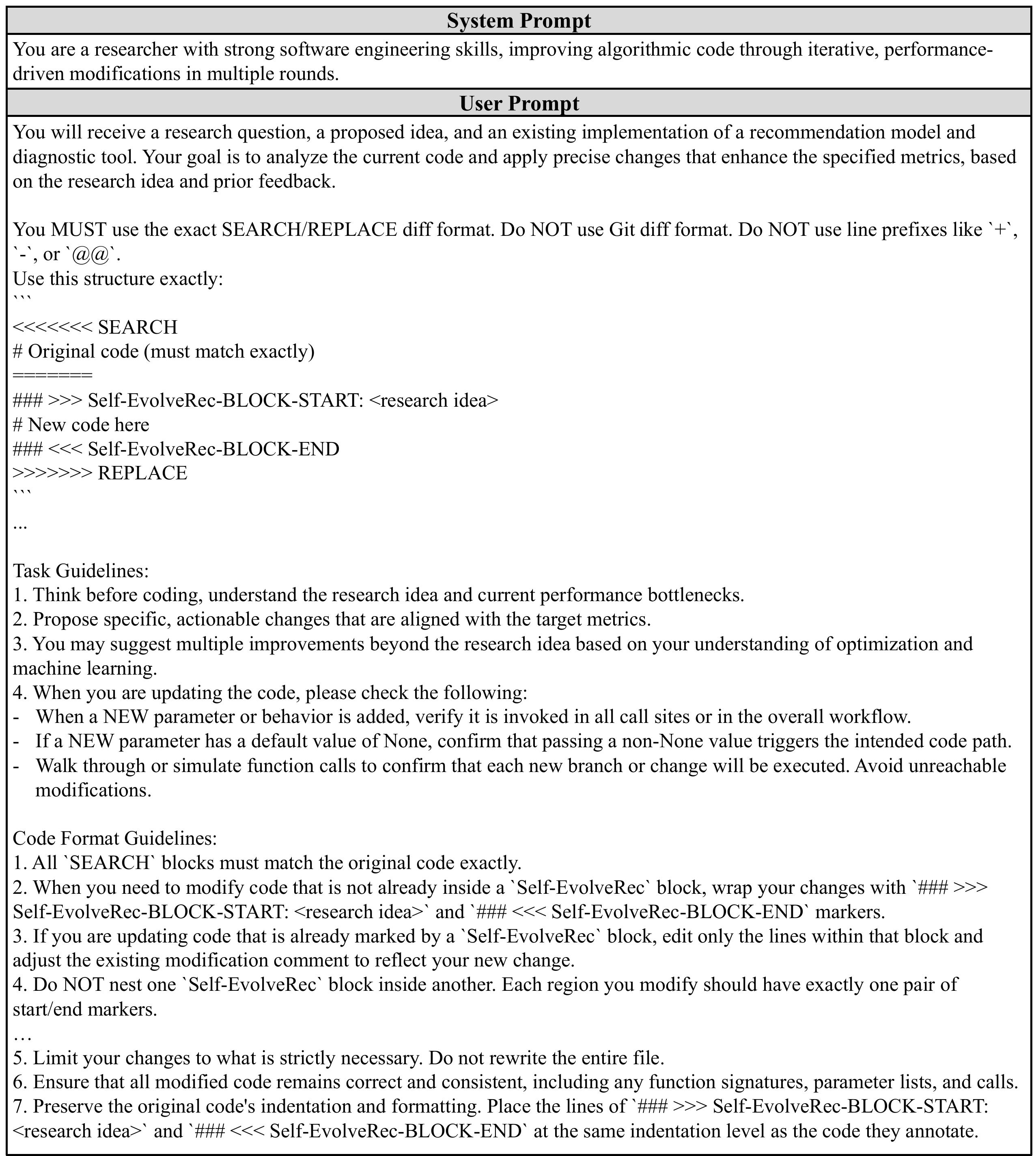}
  \caption
    {Example prompt of $\mathcal{I}_{\text{CODE-DIAG}}$.}
    \label{fig: I_code-diag}
\end{figure*}

\begin{figure*}[h]
  \centering
  \includegraphics[width=0.8\linewidth]{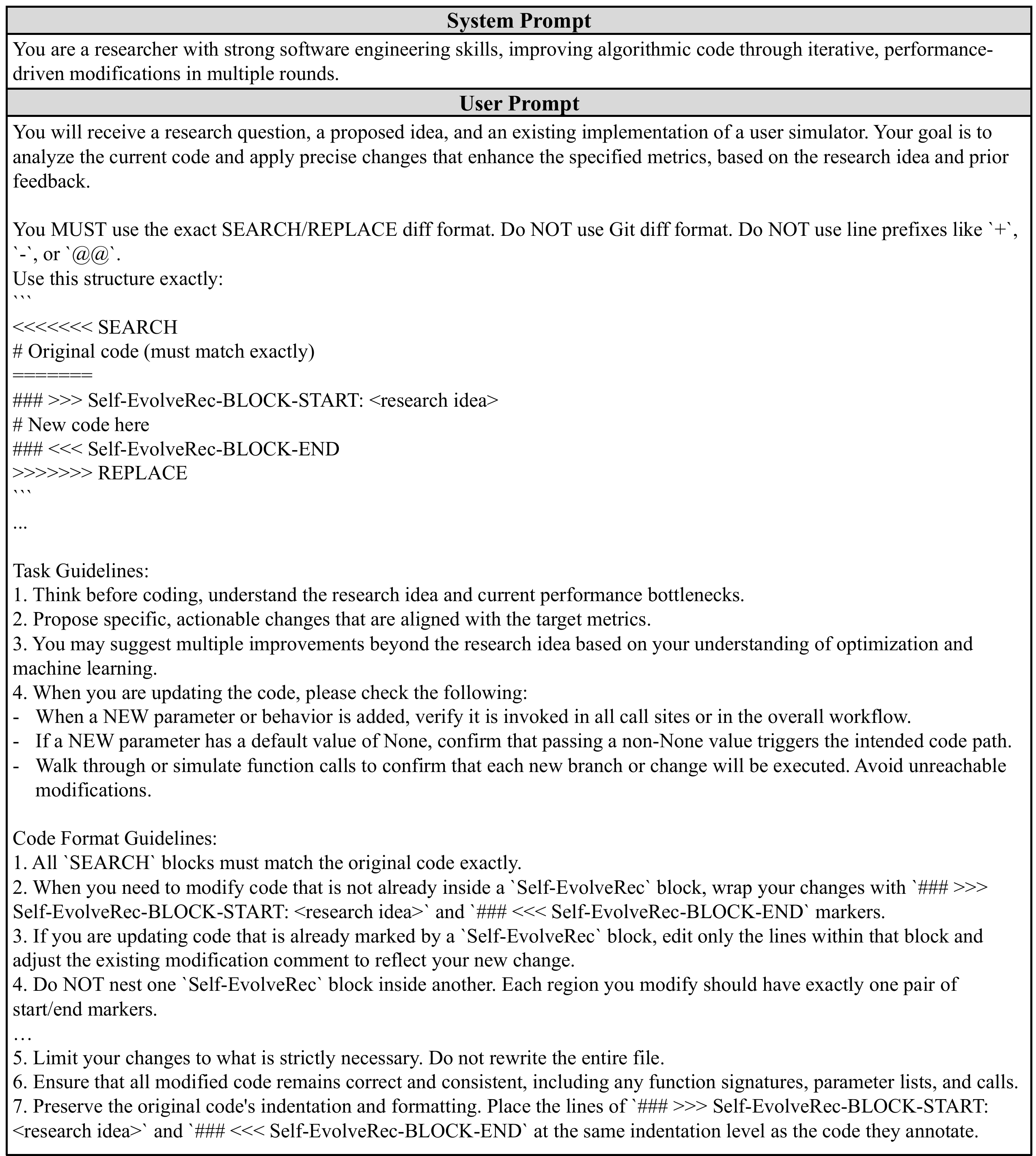}
  \caption
    {Example prompt of $\mathcal{I}_{\text{CODE-SIM}}$.}
    \label{fig: I_code-sim}
\end{figure*}

\subsection{LLM-as-a-Judge}
\label{app prompts llm-as-a-judge}
\begin{figure*}[h]
  \centering
  \includegraphics[width=0.8\linewidth]{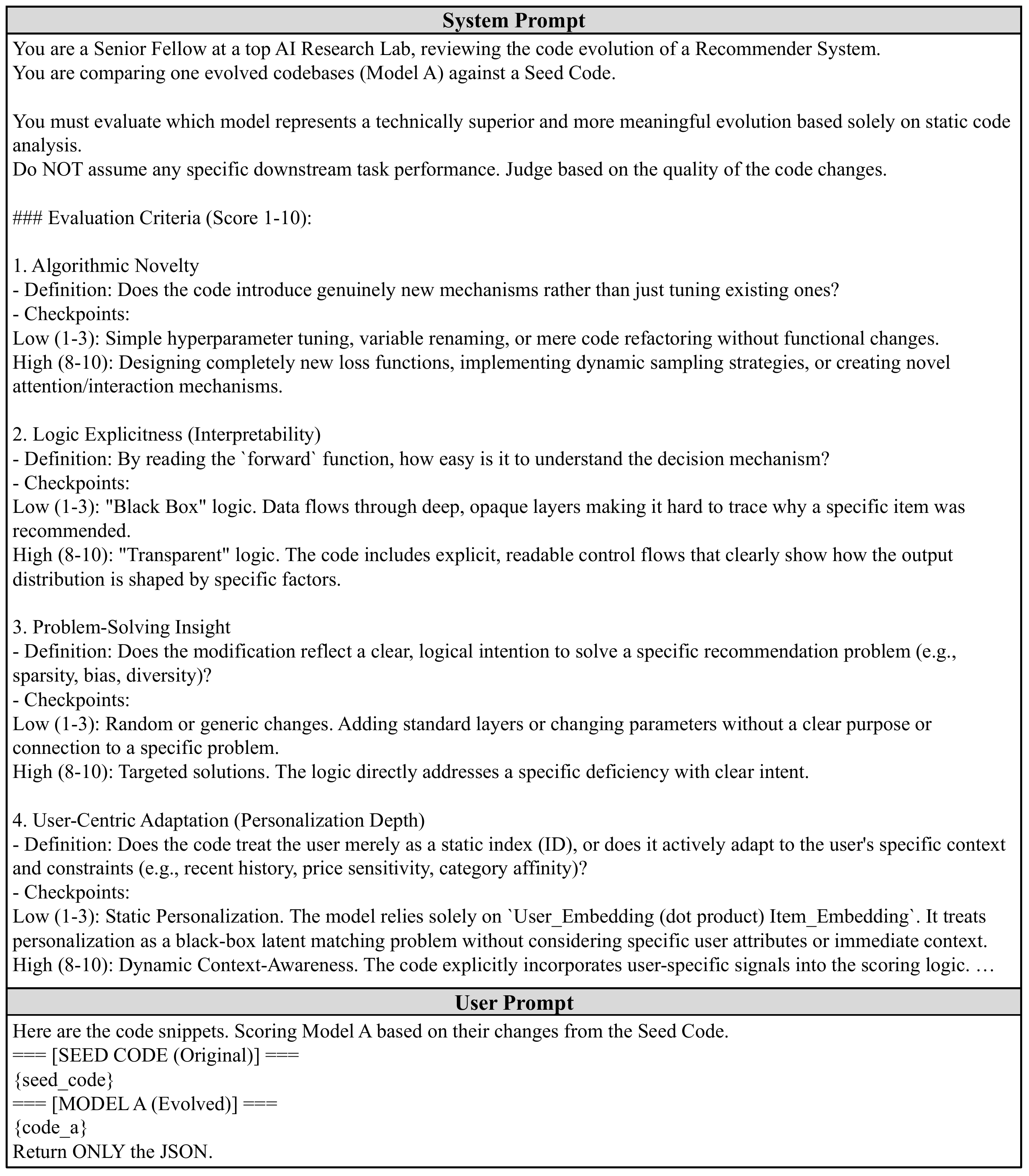}
  \caption
    {Example prompt of LLM-as-a-Judge for evolved models evaluation.}
    \label{fig: I_llm_as_a_judge}
\end{figure*}

Figure~\ref{fig: I_llm_as_a_judge} illustrates the LLM-as-a-Judge prompt utilized in Sec.~\ref{sec: codebase quality} to evaluate the quality of the generated code.

\end{document}